\pdfoutput=1

\documentclass[11pt,twoside,a4paper,cmspaper,final,collab]{cms-tdr}
\def\svnVersion{256402}\def\cmsCernNo{2014-073}\def\cmsCernDate{\today}\def\cmsMessage{Published in Physics Letters B as \href{http://dx.doi.org/10.1016/j.physletb.2014.07.053}{\doi{10.1016/j.physletb.2014.07.053.}}}

\begin{document}\cmsNoteHeader{SUS-13-024}

\hyphenation{had-ron-i-za-tion}
\hyphenation{cal-or-i-me-ter}
\hyphenation{de-vices}

\RCS$Revision: 255120 $
\RCS$HeadURL: svn+ssh://svn.cern.ch/reps/tdr2/papers/SUS-13-024/trunk/SUS-13-024.tex $
\RCS$Id: SUS-13-024.tex 255120 2014-08-07 20:09:29Z alverson $
\newlength\cmsFigWidth
\ifthenelse{\boolean{cms@external}}{\setlength\cmsFigWidth{0.98\columnwidth}}{\setlength\cmsFigWidth{0.6\textwidth}}
\ifthenelse{\boolean{cms@external}}{\providecommand{\cmsLeft}{top}}{\providecommand{\cmsLeft}{left}}
\ifthenelse{\boolean{cms@external}}{\providecommand{\cmsRight}{bottom}}{\providecommand{\cmsRight}{right}}
\newcommand{\lumin}{19.5\fbinv}
\newcommand{\Ttt}{\ensuremath{\PSQt \to \cPqt \PSGczDo}\xspace}
\newcommand{\TbW}{\ensuremath{\PSQt \to \cPqb \PSGcp}\xspace}
\newcommand{\ttll}{\ensuremath{\ttbar\to\ell\ell}\xspace}
\newcommand{\ttlj}{\ensuremath{\ttbar\to\ell+\text{jets}}\xspace}
\newcommand{\ttsl}{\ensuremath{\ttbar\to\ell+\text{jets}}\xspace}
\newcommand{\ttdl}{\ensuremath{\ttbar\to\ell\ell+\text{jets}}\xspace}
\newcommand{\ttbarW}{\ensuremath{\ttbar\PW}\xspace} %
\newcommand{\ttbarZ}{\ensuremath{\ttbar\cPZ}\xspace} %
\newcommand{\ttbarg}{\ensuremath{\ttbar\PGg}\xspace} %
\newcommand{\ttjets}{\ensuremath{\ttbar+\text{jets}}\xspace}
\newcommand{\wjets}{\ensuremath{\PW+\text{jets}}\xspace}
\newcommand{\wwjets}{\ensuremath{\PW\PW+\text{jets}}\xspace}
\newcommand{\wzjets}{\ensuremath{\PW\cPZ+\text{jets}}\xspace}
\newcommand{\zzjets}{\ensuremath{\cPZ\cPZ+\text{jets}}\xspace}
\newcommand{\zjets}{\ensuremath{\cPZ+\text{jets}}\xspace}
\providecommand{\mt}{\ensuremath{m_{\mathrm{T}}}\xspace}
\newcommand{\mtw}{\ensuremath{m_{\mathrm{T}2}^{\PW}}\xspace}
\newcommand{\lsp}{\PSGczDo\xspace}
\newcommand{\chipo}{\ensuremath{\chipm_1}}
\newcommand{\Lep}{\ensuremath{\ell}\xspace}
\newcommand{\htratio}{\ensuremath{H_{\mathrm{T}}^{\text{ratio}}}\xspace}
\newcommand{\nbb}{\ensuremath{N_{\cPqb\cPqb}}\xspace}
\newcommand{\mbb}{\ensuremath{m_{\cPqb\cPqb}}\xspace}
\newcommand{\nbjets}{\ensuremath{N_{\cPqb~\text{jets}}}\xspace}
\newcommand{\njets}{\ensuremath{N_{\text{jets}}}\xspace}
\newcommand{\bb}{\ensuremath{\cPqb\cPqb}\xspace}
\newcommand{\bjet}{\ensuremath{\cPqb{\text{ jet}}}\xspace}
\newcommand{\bjets}{\ensuremath{\cPqb{\text{ jets}}}\xspace}
\newcommand{\bhjet}{\ensuremath{\cPqb{\text{-jet}}}\xspace}
\newcommand{\mindphi}{\ensuremath{\min \Delta \phi}}
\newcommand{\qqbar}{\ensuremath{\cPq\cPaq}\xspace}
\providecommand{\MT}{\ensuremath{m_{\mathrm{T}}}\xspace}
\providecommand{\FASTJET}{\textsc{fastjet}\xspace}

\newcommand\tablespace{\rule[-1ex]{0cm}{3.5ex}}

\cmsNoteHeader{SUS-13-024} 
\title{Search for top-squark pairs decaying into Higgs or Z bosons in pp collisions at $\sqrt{s}=8$\TeV}

\date{\today}

\abstract{
  A search for supersymmetry through the direct pair production of top
  squarks, with Higgs (H) or Z bosons in the decay chain, is performed
  using a data sample of proton-proton collisions at $\sqrt{s}=8$\TeV
  collected in 2012 with the CMS detector at the LHC. The sample
  corresponds to an integrated luminosity of 19.5\fbinv.
  The search is performed using a selection of events containing
  leptons and bottom-quark jets. No evidence for a significant excess of events over the standard model
  background prediction is observed.
  The results are interpreted in the context of simplified
  supersymmetric models
  with pair production of a heavier top-squark mass eigenstate $\PSQtDt$
  decaying to a lighter top-squark mass eigenstate
  $\PSQtDo$ via either
  $\PSQtDt\to \PH
  \PSQtDo$ or $\PSQtDt\to
  \cPZ\PSQtDo$, followed in both cases by
  $\PSQtDo\to \cPqt\PSGczDo$,
  where $\PSGczDo$ is an undetected, stable, lightest supersymmetric particle.
  The interpretation is performed in the region where the mass
  difference between the $\PSQtDo$ and
  $\PSGczDo$ states is approximately equal to the top-quark mass
  ($m_{\PSQtDo} - m_{\PSGczDo} \simeq
  m_{\cPqt}$), which is not probed by searches for direct $\PSQtDo$ squark pair production.
  The analysis excludes top squarks with masses $m_{\PSQtDt}<575$\GeV and
  $m_{\PSQtDo}<400$\GeV at a 95\% confidence level.
}

\hypersetup{%
pdfauthor={CMS Collaboration},%
pdftitle={Search for top-squark pairs decaying into Higgs or Z bosons in pp collisions at sqrt(s) = 8 TeV},%
pdfsubject={CMS},%
pdfkeywords={CMS, SUSY, stop, Higgs}}

\maketitle

\section{Introduction}
\label{sec:introduction}

Supersymmetry (SUSY) with R-parity conservation~\cite{Farrar:1978xj} is an extension to the standard model (SM)
that provides a candidate particle for dark matter and addresses the hierarchy problem~\cite{SUSY1,SUSY2,SUSY3,SUSY4,SUSY5,SUSY6}.
The hierarchy problem originates in the spin-zero nature of the Higgs ($\PH$) boson, whose mass is subject to
divergences from higher-order corrections. The leading
divergent contribution from SM particles arises from the $\PH$ boson
coupling to the top quark.
SUSY provides a possible means to stabilize the $\PH$ boson mass calculation,
through the addition of contributions from a scalar top quark (top-squark) with a mass not
too different from that of the top quark~\cite{Barbieri:1987fn,deCarlos1993320,Dimopoulos1995573,Barbieri199676,Papucci:2011wy}.
Searches for direct top-squark production from the ATLAS~\cite{ATLAS1,ATLAS2,ATLAS3,ATLAS4,ATLAS6,ATLAS8} and
Compact Muon Solenoid (CMS)~\cite{SUS-13-011}
Collaborations at the Large Hadron Collider (LHC) at CERN have focused mainly
on the simplest scenario, in which only the lighter top-squark
mass eigenstate, $\stone$, is accessible at current LHC collision energies. In these searches, the top-squark decay modes
considered are those to a top quark and a neutralino,
$\stone\to \cPqt\PSGczDo \to \cPqb \PW \PSGczDo$, or
to a bottom quark and a chargino, $\stone\to \cPqb \PSGcpDo \to \cPqb \PW \PSGczDo$.
These two decay modes are expected to have large branching fractions
if kinematically allowed.
The lightest neutralino, $\PSGczDo$, is the lightest
SUSY particle (LSP) in the R-parity conserving models
considered; the experimental signature of such a particle is
missing transverse energy ($\MET$).

Searches for top-squark pair production are challenging because
the cross section is approximately
six times smaller than that for top-antitop quark pair ($\ttbar$) production if $m_{\stone} \sim m_{\cPqt}$
and decreases rapidly with increasing top-squark mass~\cite{Han:2008gy}. When the mass
difference between the top-squark and the
$\PSGczDo$ is large, top-squark production can be distinguished from $\ttbar$
production, as the former is typically characterized by events with extreme kinematic features, especially large
$\MET$. This strategy is being pursued in existing searches and has
sensitivity to top-squark masses up to
about 650\GeV for low $\PSGczDo$ masses~\cite{ATLAS1,ATLAS2,ATLAS3,ATLAS4,ATLAS6,ATLAS8,SUS-13-011}.
The sensitivity of searches for direct top-squark pair production is, however, significantly reduced
in the $\stone\to \cPqt\PSGczDo$ decay mode for the region of SUSY parameter space in which
 $m_{\stone} - m_{\PSGczDo} \simeq m_{\cPqt}$.  For example, in Ref.~\cite{SUS-13-011},
the region $|m_{\stone} - m_{\PSGczDo} - m_{\cPqt}| \lesssim 20$\GeV is unexplored.
In this region, the momentum of the daughter neutralino in the rest frame of the decaying
$\stone$ is small, and it is exactly zero in the limit $ m_{\stone} - m_{\PSGczDo} = m_{\cPqt}$. As a result, the $\MET$
from the vector sum of the transverse momenta of the two neutralinos is typically also small in the laboratory frame.
It then becomes difficult to distinguish kinematically between
$\stone$ pair production and the dominant background,
which arises from $\ttbar$ production.
This region of phase space can be explored using events with topologies that are distinct from the $\ttbar$ background.
An example is gluino pair production where each gluino decays to a top squark and a top quark, giving rise to a
signature with four top quarks in the final state~\cite{Chatrchyan:2013iqa,Chatrchyan:2013wxa}.

\begin{figure*}[bth]
  \centering
    \includegraphics[width=0.33\linewidth]{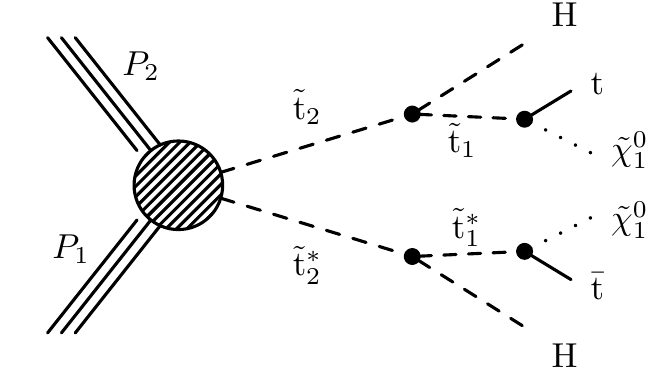}%
    \includegraphics[width=0.33\linewidth]{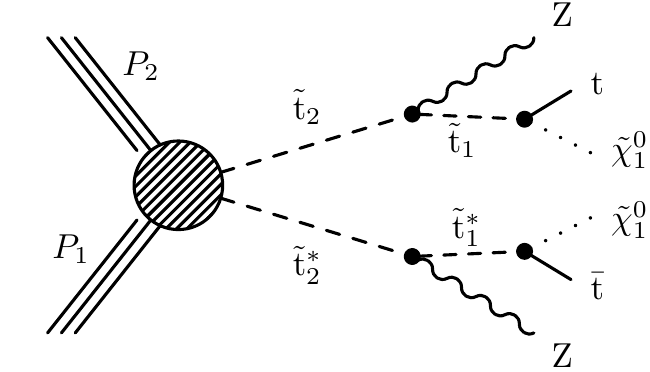}%
    \includegraphics[width=0.33\linewidth]{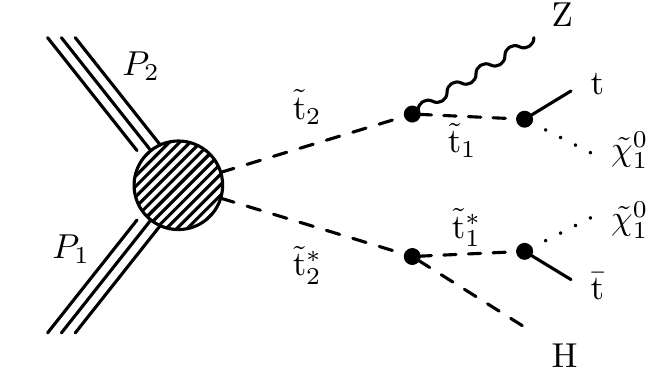}
    \caption{
    \label{fig:T6ttXXdiagram}
    Diagrams for the production of the heavier top-squark ($\sttwo$)
    pairs followed by the decays $\sttwo\to \PH \stone$ or $\sttwo\to
  \cPZ \stone$ with $\stone \to \cPqt \PSGczDo$.
  The symbol * denotes charge conjugation.}
\end{figure*}

This analysis targets the region of phase space where
$m_{\stone} - m_{\PSGczDo} \simeq m_{\cPqt}$ by focusing on signatures
of $\ttbar\PH\PH$, $\ttbar\PH\cPZ$, and
$\ttbar\cPZ\cPZ$ with $\MET$. These final states can
arise from the pair production of the heavier top-squark mass eigenstate
$\sttwo$. There are two non-degenerate top-squark mass eigenstates
($\sttwo$ and $\stone$) due to the mixing of the SUSY partners $\stL$ and $\stR$ of the
right- and left-handed top quarks. The $\sttwo$
decays to $\stone$ and an $\PH$ or $\cPZ$ boson,
and the $\stone$ is subsequently assumed to decay to $\cPqt\PSGczDo$, as
shown in Fig.~\ref{fig:T6ttXXdiagram}. Other decay modes such as $\stone\to
\cPqb\PSGcpDo \to \cPqb\PW\PSGczDo $ are largely covered
for $m_{\stone} - m_{\PSGczDo} \simeq m_{\cPqt}$ by existing analyses~\cite{SUS-13-011}.
The final states pursued in this search can arise in other scenarios, such as $\stone\to
\cPqt\PSGczDt$, with $\PSGczDt \to \PH
\PSGczDo$ or $\PSGczDt \to \cPZ
\PSGczDo$. The analysis is also sensitive to a range of models in
which the LSP is a gravitino~\cite{Aad:2012cz,ATLAS7}.
The relative branching fractions for modes with the $\PH$ and $\cPZ$ bosons are
model dependent, so it is useful to search for both decay modes simultaneously.
In the signal model considered, $\sttwo$ is assumed always to decay to
$\stone$ in association with an $\PH$ or $\cPZ$ boson,
such that the sum of the two branching fractions is
$\mathcal{B}(\sttwo\to \PH \stone) + \mathcal{B}(\sttwo\to \cPZ
\stone) = 100$\%. Other possible decay modes are $\sttwo \to
\cPqt\PSGczDo$ and $\sttwo \to \cPqb \PSGcpDo$. These
alternative decay modes are not considered here, since they
give rise to final states that are covered by existing searches for
direct top-squark pair production~\cite{ATLAS1,ATLAS2,ATLAS3,ATLAS4,ATLAS6,ATLAS8,SUS-13-011}.

The results are based on proton-proton collision data collected at
$\sqrt{s}=8$\TeV by the CMS experiment at the LHC during
2012, corresponding to an integrated luminosity of $\lumin$.
The analysis presented here searches for $\sttwo$ production in a sample of events with
charged leptons, denoted by $\ell$ (electrons or muons), and jets identified as originating from bottom quarks
($\cPqb$ jets). The four main search channels contain either exactly
one lepton, two leptons with opposite-sign (OS) charge and no other
leptons, two leptons with same-sign (SS) charge and no other leptons,
or at least three leptons (3 $\ell$).
The channels with one lepton or two OS leptons require at least three
$\cPqb$ jets, while the channels with two SS leptons or 3 $\ell$ require
 at least one $\cPqb$ jet.
These requirements suppress background contributions from $\ttbar$ pair production,
which has two $\cPqb$ quarks and either one lepton or two OS leptons from
the $\ttbar\to \ell \nu \qqbar \bbbar$ or
$\ttbar\to \ell \nu \ell \nu \bbbar$ decay modes, where
$\cPq$ denotes a quark jet. The
sensitivity to the signal arises both from events with additional
$\cPqb$ quarks in the final state (mainly from
$\PH\to\bbbar$),
and from events with additional leptons from $\PH$ or $\cPZ$ boson decays.

This letter is organized as follows: Section~\ref{sec:detector} briefly
introduces the CMS detector, while Section~\ref{sec:samples} presents
the event samples and the object selections
used. Section~\ref{sec:sigregions} describes the signal regions, and
Section~\ref{sec:backgrounds} details the background estimation
methods. The experimental results are presented in
Section~\ref{sec:results}, and in Section~\ref{sec:interpretation} we discuss the interpretation of the results
in the context of the signal model of the pair production
of a heavier top-squark mass eigenstate $\sttwo$ decaying to a
lighter top-squark mass eigenstate $\stone$.

\section{The CMS detector}
\label{sec:detector}

The CMS detector~\cite{JINST} comprises a silicon tracker
surrounded by a lead-tungstate crystal electromagnetic calorimeter
and a brass-scintillator hadronic calorimeter, a superconducting
solenoid
supplying a 3.8\unit{T} magnetic field to the detectors enclosed, and a muon system.
The silicon tracker system consists of pixel and strip detectors, which
measure the trajectories of charged particles. Energy measurements of
electrons, photons, and hadronic jets are provided by the
electromagnetic and hadronic calorimeters. Each of these systems
includes both central (barrel) and forward (endcap) subsystems.
These detectors operate in the axial magnetic field of the solenoid,
while muons are identified in gas-ionization detectors that are
embedded in the steel flux-return yoke of the solenoid.

The CMS experiment uses a right-handed
coordinate system with the origin at the
nominal pp interaction point at the center of the detector.
The positive $x$ axis is defined by the direction
from the interaction point to the center of the LHC ring, with the
positive $y$ axis pointing upwards. The azimuthal angle $\phi$ is
measured around the beam axis in radians and the polar angle $\theta$
is measured from the $z$ axis pointing in the direction of the counterclockwise
LHC beam. The pseudorapidity is defined as $\eta \equiv -\ln [\tan (\theta/2)]$.

The silicon tracker, the muon system, and the electromagnetic calorimeter
cover the regions $\abs{\eta}<2.4$, $\abs{\eta}<2.4$, and  $\abs{\eta}<2.5$,
respectively.
The hadronic calorimeters extend up to $\abs{\eta} \approx 5$,
improving momentum balance
measurements in the plane transverse to the beam direction.
The online trigger system that selects collision events of interest is based on
two stages: a first-level hardware-based selection and a second set
of requirements implemented in software.

\section{Event samples, object selection, and event simulation}
\label{sec:samples}

The data used for this search were collected with a high transverse-momentum ($\pt$) electron ($\Pe$) or muon ($\mu$) single-lepton trigger, which requires
at least one electron with $\pt>27\GeV$ or muon
with $\pt>24\GeV$. The trigger efficiencies, as measured with a sample of
$\cPZ\to\ell^+\ell^-$ events, vary between 85\% and 97\% for
electrons, and between 80\% and 95\% for muons, depending on the
$\eta$ and $\pt$ values of the leptons.
Events were also collected with the $\Pe\Pe$, $\Pe\mu$, and $\mu\mu$
double-lepton triggers, which require at least one $\Pe$ or one $\mu$
with $\pt>17\GeV$ and another with $\pt>8\GeV$.
Events are also acquired with a double-lepton trigger targeting
lower-$\pt$ leptons, requiring $\pt>8\GeV$, but with an additional
online selection of $\HT\equiv\Sigma_\text{jet} |\pt^\text{jet}| > 175$\GeV,
considering only jets with $\pt>40\GeV$ in the sum.
The efficiencies lie between 90\% and 95\% for the trigger targeting
lower-$\pt$ leptons, and between 80\% and 95\% for the trigger
targeting higher-$\pt$ leptons, depending on the
$\eta$ and $\pt$ values of the lower-$\pt$ lepton. For selections with
more than two leptons, the triggers are fully efficient.

Events are reconstructed offline using the particle-flow (PF)
algorithm~\cite{PFT-09-001,PFT-10-001}.
Electron candidates are reconstructed by associating tracks
with energy clusters in the electromagnetic
calorimeter~\cite{EGM-10-004,Chatrchyan:2013dga}. Muon candidates are reconstructed by
combining information from the tracker and the muon
detectors~\cite{MUO-10-004}.
Signal leptons are produced in the decays of $\PW$ and $\cPZ$ bosons.
In order to distinguish these leptons from those produced
in the decays of heavy-flavor hadrons,
all lepton candidates are required to be consistent with originating
from the primary interaction vertex, chosen as the vertex with the
highest sum of the $\pt^2$ of its constituent tracks. In particular
they are required to have a transverse impact parameter with respect to this
vertex smaller than 0.2\unit{mm}. A tighter requirement is used for the
event category with two SS leptons (see
Ref.~\cite{CMS-PAS-SUS-13-013}).
Furthermore, since misidentified lepton candidates arising
from background sources, such as the decays of hadrons, are
typically embedded in jets, all lepton candidates are required to be
isolated from hadronic activity in the event. This is achieved by
imposing a maximum allowed value on the quantity $\pt^{\text{sum}}$,
defined as the scalar sum of the $\pt$ values of charged and neutral
hadrons and photons within a cone of radius $\Delta R \equiv \sqrt{\smash[b]{(\Delta\eta)^2+(\Delta\phi)^2}}=0.3$
around the lepton candidate momentum direction at the origin.
For the event category with at least three leptons, the isolation
requirement is $\pt^{\text{sum}} < 0.15 \pt$. For the lower
lepton-multiplicity selections, the isolation requirement is tighter
(see Refs.~\cite{SUS-13-011} and~\cite{CMS-PAS-SUS-13-013} for details).
The surrounding hadronic activity is corrected for the energy
contribution from additional proton-proton interactions in
the event (pileup), as described in Ref.~\cite{HIG-13-004}.

Jets are reconstructed from particle-flow candidates using the anti-\kt clustering algorithm~\cite{Cacciari:2008gp}
with a distance parameter of 0.5. Their energies are corrected for residual
non-uniformity and non-linearity of the detector response using
corrections derived from exclusive dijet and $\gamma/\cPZ$+jet data~\cite{Chatrchyan:2011ds}.
The energy contribution from pileup is estimated using the jet
area method for each event~\cite{cacciari-2008-659} and is subtracted
from the jet $\pt$. Only high-$\pt$ jets in the central
calorimeter $\abs{\eta}<2.4$ are considered. Jets consistent with the
decay of heavy-flavor hadrons are identified using the combined
secondary vertex b-tagging algorithm at the medium or loose working
points, defined such that they have tagging efficiencies of 70\% or 80--85\%, and
misidentification rates for light-flavor jets less than 2\% or
10\%, respectively~\cite{ref:btag}.
The $\MET$ is calculated as the magnitude of
the vector sum of the transverse momenta of
all PF candidates, incorporating jet energy corrections~\cite{CMS-PAS-JME-12-002}.
Quality requirements are applied to
remove a small fraction of events in which detector effects such as electronic noise
can affect the $\MET$ reconstruction.
Events are required to have $\MET>50\GeV$ to reduce
background contributions from sources with a single $\PW$ boson and
from jet production via QCD processes.

Simulated event samples are used to study the characteristics of the
signal and to calculate its acceptance, as well as for part of the
SM background estimation. Pair production of $\sttwo$ squarks is
described by the \MADGRAPH 5.1.3.30~\cite{madgraph5} program,  including up to two
additional partons at the matrix element level, which are matched to the
parton showering from the \PYTHIA 6.424~\cite{Sjostrand:2006za} program. The SUSY particle decays are
simulated with \PYTHIA with a uniform amplitude over phase
space, so that all decays are isotropic~\cite{Chatrchyan:2013sza}.
The first two decay modes considered (see Fig.~\ref{fig:T6ttXXdiagram}) are assumed to have a branching
fraction of unity when setting limits on SUSY particle masses. The Higgs boson mass
is set to 125\GeV~\cite{CMS_Higgs}, and its branching fractions are set according to the
corresponding expectations from the SM~\cite{Dittmaier:2012vm}.
For each decay mode, a grid of signal events is generated as a function
of the two top-squark masses $m_{\sttwo}$ and
$m_{\stone}$. The $\stone$ squark is forced to decay to a top quark and a
neutralino LSP assuming $m_{\stone} - m_{\PSGczDo} = 175$\GeV. The
top-quark mass is set to 175\GeV. The signal event rates are normalized to cross sections
calculated at next-to-leading order (NLO) in the strong coupling
constant, including the resummation of soft-gluon emission at
next-to-leading-logarithmic accuracy
(NLO+NLL)~\cite{bib-nlo-nll-01,bib-nlo-nll-02,bib-nlo-nll-03,bib-nlo-nll-04,bib-nlo-nll-05,ref:xsec}.

The SM background processes considered are the production of $\ttbar$;
\ttbar\ in association with a boson ($\PH$, $\PW$, $\cPZ$, $\gamma^{*}$);
$\PW$, $\cPZ$, and $\gamma^{*}$+jets; triboson; diboson; single-top
quark in the $s$, $t$, and $\cPqt\PW$ channels; and single-top
quark in association with an additional
quark and a $\cPZ$ boson. These processes are generated with the
\MADGRAPH, \POWHEG-box 1.0~\cite{Alioli:2009je,Re:2010bp}, or \MCATNLO 2.0.0 beta3~\cite{MCatNLO1,MCatNLO2} programs,
using the CT10~\cite{ct10} (\POWHEG), CTEQ6M~\cite{cteq6lm}
(\MCATNLO), and CTEQ6L1~\cite{cteq6lm} (\MADGRAPH)
parton distribution functions (PDFs). SM background event rates are
normalized to cross
sections~\cite{MCatNLO1,MCatNLO2,xsec_ttbar,xsec_ttbarW,xsec_ttbarZ,xsec_MCFM,xsec_WZ,MCatNLO3}
calculated at next-to-next-to-leading
order when available, otherwise at NLO.
All the background samples are processed with the full simulation of the CMS detector
based on \GEANTfour~\cite{Geant}, while the generated signal samples
use a fast simulation~\cite{Abdullin:2011zz}. The fast simulation is
validated against the full simulation for the variables relevant
for this search, and efficiency corrections based on data are applied~\cite{Rahmat:2012fs}.
The simulation is generated with inelastic collisions superimposed on
the hard-scattering event. Events are weighted so that the distribution of the number
of inelastic collisions per bunch crossing matches that in data.

\section{Event categories and signal regions}
\label{sec:sigregions}

The search is carried out through comparisons of the data and SM
background yields in disjoint signal regions (SRs) targeting the SUSY processes shown in
Fig.~\ref{fig:T6ttXXdiagram}, while suppressing the contributions from SM
backgrounds, predominantly $\ttbar$ production. The definitions of the
SRs are summarized in
Table~\ref{tab:SummarySelection}, and are detailed in the following subsections.
Events are classified according to the
lepton multiplicity and charge requirements on the leptons. Four main
event categories are considered. The first two include events with one
lepton or two OS leptons. Since these lepton signatures also arise in
the decays of top-antitop quark pairs, requirements of at least three \bjets\ are used to
suppress this background. The other two categories are events
with exactly two SS leptons and events with three or more leptons,
which do not typically arise in $\ttbar$ events.
A requirement of at least one \bjet\ is applied to further suppress the
contribution from backgrounds from $\PW$ and $\cPZ$
bosons. Lepton vetoes are used to ensure that the four main event
categories do not overlap.

\begin{table*}[tbh]
\centering
\topcaption{Summary of the SR definitions for the different
  selections, specified by rows in the table. The SRs correspond to all possible
  combinations of requirements in each row, where different regions for the kinematic
  variables are separated by commas. For the event category with two SS leptons, two
  selections in lepton $\pt$ are used (low and high), as explained in the
  text. There are 96 SRs in total.\label{tab:SummarySelection}}
\resizebox{\textwidth}{!}{
\begin{tabular}{l c c c c c}
\hline
$N_{\ell}$          & Veto & $\nbjets$ & $\njets$ & $\ETmiss$ [\GeVns{}] & Additional requirements [\GeVns{}] \tablespace \\
\hline\hline
\multirow{2}{*}{1}  		  & \multirow{2}{*}{track or $\tau_\mathrm{h}$} & $=$3    &  $\ge$5 & \multirow{2}{*}{$\ge$50}                  & $\MT>150$  \\
&  & $\ge$4              &  $\ge$4 &   & $\MT>120$ \\\cline{1-6}
\multirow{2}{*}{2 OS} 	  & \multirow{2}{*}{extra  $\Pe$/$\mu$}
& $=3$    & $\ge$5  &  \multirow{2}{*}{$\ge$50}   &
\multirow{2}{*}{$\nbb=1$ with $100\le \mbb\le150$ or $\nbb\ge2$}\\
& & $\ge$4  &  $\ge$4  &   &    \\\hline 
\multirow{2}{*}{2 SS}  & \multirow{2}{*}{extra  $\Pe$/$\mu$} & $=$1   &
\multirow{2}{*}{$[2,3],\ge 4$} & \multirow{2}{*}{[$50,120$], $\ge$120} &
\multirow{2}{*}{for low (high) \pt: $250(200)\le\HT\le400$, $\HT\ge$400}  \\
& & $\ge$2  &  &  &  \\\hline 
\multirow{3}{*}{$\ge$3}  & \multirow{3}{*}{---}  & $=1$  &
\multirow{2}{*}{$[2, 3],\ge4$} &  &
\multirow{3}{*}{for on/off-$\cPZ$: $60\le\HT\le200$, $\HT\ge$200}\\
 &  &  $=2$ &  &  [$50,100$], [$100,200$], $\ge$200 & \\\cline{3-4}
 &  &  $\ge$3 & $\ge$3 &   & \\\hline
\end{tabular}
}
\end{table*}

\subsection{Event categories with a single lepton or two opposite-sign leptons}
\label{sec:1l2OSlSelection}
The event categories with one lepton or two OS leptons, accompanied
in either case by at least three \bjets, target signatures with $\PH$
bosons, which have large branching fraction for $\PH \to \bbbar$.
In the single-lepton channel, events are required to have exactly one
electron with $\pt>30$\GeV and $\abs{\eta}<1.44$ or exactly one muon with
$\pt>25$\GeV and $\abs{\eta}<2.1$. Events with an indication of an
additional lepton, either an isolated track~\cite{CMS-PAS-SUS-13-013} or a
hadronically decaying $\tau$-lepton candidate $\tau_\mathrm{h}$~\cite{PFT-08-001,PFT-10-004,CMS-PAS-TAU-11-001},
are rejected in order to reduce the background from \ttbar events
in which both W bosons decay leptonically.
In the double-lepton channel, events are required to contain exactly two charged leptons ($\Pe\Pe$, $\Pe\mu$, or $\mu\mu$),
each with $\pt>20$\GeV and $\abs{\eta}<2.4$. In this case,
events with an additional $\Pe$ or $\mu$ with $\pt> 10$\GeV are
rejected. Any electron candidate in the region $1.44<\abs{\eta}<1.57$,
a less well-instrumented transition region between the barrel and
endcap regions of the calorimeter,  is excluded in the event selection
since standard electron identification capabilities are not optimal.
Jets are required to be separated from the candidate leptons by $\Delta R > 0.4$.

In these event categories, a typical $\ttbar$ background event has two \bjets\
in the final state, while signal events could have up to four
additional \bjets, two from each $\PH$ decay. The requirement of more than two \bjets\ greatly
suppresses the \ttbar\ background contribution.
For events with exactly three $\bjets$, the jet $\pt$ threshold applied is
40\GeV; for events with at least four $\bjets$, the threshold is
lowered to 30\GeV. In both cases, the medium working point of the
$\bhjet$ tagger is used (see Section~\ref{sec:samples}).
To further reduce the \ttbar\ background contribution in the sample
with exactly three \bjets, events are required to contain two additional jets with
$\pt>30\GeV$, at least one of which must satisfy the loose but not the medium criteria of the b-jet tagger.
Signal 
events can have large jet and $\bhjet$ multiplicities,
while in the case of the \ttbar\ background, additional jets are
needed to satisfy this selection criterion. To reduce the
contribution of jets from pileup in
the event, a requirement is applied on a multivariate discriminating variable that incorporates
the multiplicity of objects clustered in the jet, the jet shape, and
the compatibility of the charged constituents of the jet with the
primary interaction vertex~\cite{CMS-PAS-JME-13-005}.

Besides the requirements listed above, the analysis in the single-lepton channel selects events
with large transverse mass of the $(\ell,\nu)$ system,
defined as $\mt\equiv\sqrt{\smash[b]{2\pt^{\ell}\pt^{\nu} [1-\cos
  (\phi^{\ell}-\phi^{\nu})]}}$, where the $\pt$ of the selected lepton is used and the $(x,y)$
components of the neutrino momentum are equated to the
corresponding $\MET$ components. For events in which the $\MET$ arises
from a single neutrino from a $\PW$ boson decay, this variable
has a kinematic endpoint $\mt \approx m_{\PW}$, where
$m_{\PW}$ is the $\PW$ boson mass.
The requirement of large \mt\ ($\mt>150$\GeV for events with three
\bjets\ or $\mt>120$\GeV for events with at least four \bjets) provides strong suppression
of the semileptonic \ttbar\ background.

The study of the OS dilepton channel uses information from pairs
of \bjets\ (ignoring their charge) to identify pairs consistent
with $\PH \to \bbbar$ decay:
$\Delta R_{\cPqb\cPqb} \le 2\pi/3$,
$m_{\cPqb\cPqb}$/$[\pt^{\cPqb\cPqb}\Delta R_{\cPqb\cPqb}] \le 0.65$, and
$\abs{\Delta y_{\cPqb\cPqb}}\le 1.2$, where the rapidity is defined as $y\equiv
\frac{1}{2}\ln[({E}+p_{z})/({E}-p_{z})]$,
with $p_{z}$ denoting the component of the momentum along the beam
axis. Only \bjets\ satisfying the medium working point of the tagger are
used to form $\cPqb\cPqb$ combinations. Different $\bhjet$ pairs are
not allowed to have $\bhjet$s in common.
We denote the number of selected $\bhjet$ pairs as $\nbb$ and the
invariant mass of a pair as $\mbb$. Events are required to have either
$\nbb=1$ and $100 \le \mbb \le 150$\GeV, or else $\nbb \ge 2$.
For the signal models of interest, particularly the $\sttwo\to
\PH \stone$ decay mode, the SRs with largest \bhjet\
multiplicity ($\ge$4 \bjets) have the highest sensitivity.

\subsection{Event category with two SS leptons}
The event category with two SS leptons targets signatures with
multiple sources of leptons. Standard model
processes with two SS leptons are extremely rare.
The analysis for this event category closely follows that
described in Ref.~\cite{CMS-PAS-SUS-13-013}. The only difference
is the addition of a veto on events containing a third lepton,
to remove the overlap with the 3 $\ell$ event category.
These SRs also recover events with three leptons in which one
of the three leptons falls outside the detector acceptance or fails
the selection criteria.
Multiple SRs are defined for the SS event category based on the jet
and \bhjet\ multiplicities, $\MET$, and $\HT$, and on whether the leptons
satisfy $\pt>10\GeV$ (low-$\pt$ analysis) or $\pt>20\GeV$ (high-$\pt$ analysis).
The leptons must appear within $\abs{\eta}<2.4$. The jet $\pt$
threshold applied is 40\GeV.
The low and high lepton $\pt$ samples, which partially overlap, target complementary
signatures. The low-$\pt$ sample
extends the sensitivity to signatures with compressed SUSY spectra,
while the high-$\pt$ analysis targets scenarios with leptons produced
via on-shell $\PW$ and $\cPZ$ bosons. Only the high-$\pt$ analysis
is used to target the signals explicitly studied in this letter, while the
low-$\pt$ analysis is included for sensitivity to other new physics
scenarios.

\subsection{Event category with at least three leptons}
The event category with at least three leptons and at least one \bjet\
is sensitive to all of the processes shown in
Fig.~\ref{fig:T6ttXXdiagram}. These processes contain many sources of
leptons, such as $\cPZ$ bosons from the top-squark decays, and $\tau$ leptons
and $\PW$ and $\cPZ$ bosons from the $\PH$ boson decays.
Even though signatures giving rise to three or more leptons have small production rates,
this event category has good sensitivity because the backgrounds are strongly suppressed.
The dataset is acquired using the double-lepton triggers.
Events are selected offline by requiring at least three $\Pe$ or $\mu$
candidates with $\pt > 10\GeV$, including
at least one with $\pt > 20\GeV$, and $\abs{\eta}<2.4$.
Events with two leptons of opposite-sign charge with an invariant mass
below 12\GeV are removed from the sample to reduce the contribution
of leptons originating from low-mass bound states.

Events are required to have at least two jets with
$\pt>30\GeV$ and at least one \bjet\ satisfying the
medium working point of the tagger.
Leptons within $\Delta R <0.4$ of a $\cPqb$-quark jet
are not considered isolated and are merged with the \bjet.
This requirement imposes an additional isolation criterion for leptons
and reduces the dominant background,
$\ttbar$ production, by 25--40\% depending on the SR,
compared to the case where such an object is reconstructed as a lepton
rather than a $\cPqb$ jet. The efficiency for signal leptons is
reduced by 1\%. The remaining SM background in the ${\ge}$ 3 $\ell$
event category from $\PW\cPZ$+jets
production is highly suppressed by the \bhjet\ requirement.

This three-lepton event sample is divided into several SRs
by imposing requirements on the jet and \bjet\ multiplicity, $\MET$, and
the hadronic activity in the event, as given by the kinematic
variable $\HT$, considering jets with $\pt>30\GeV$ in the sum.
Finally, events are classified as either ``on-$\cPZ$'' if there is a pair of leptons
with the same flavor and opposite charge that has an invariant mass
within 15\GeV of the nominal $\cPZ$ boson mass, or
``off-$\cPZ$'' if no such pair exists or if the invariant mass lies outside this range.

The separation of events into these SRs improves the
sensitivity of the search. For the signal models of interest,
the SRs with large \bjet\ multiplicity (those designated N$_{\bjets}=2$ and $N_{\bjets}\geq$3)
and that with both high $\MET$ and high $\HT$ provide the greatest sensitivity. The on-$\cPZ$
regions are the most sensitive, when the decay to an on-shell $\cPZ$
boson is kinematically allowed for the $\sttwo\to \cPZ \stone$
decay mode. Conversely, the off-$\cPZ$ regions have more sensitivity
when on-shell $\cPZ$ boson decays are not kinematically allowed
and for the $\sttwo\to \PH \stone$ decay mode.

\section{Background estimation}
\label{sec:backgrounds}

The main background arises from SM $\ttbar$
events, which usually have two \bjets\
and at most two leptons from $\PW$ boson decays.
Thus, $\ttbar$ events can only satisfy the
selection criteria if accompanied by sources of additional \bjets\ or
leptons. Such backgrounds are estimated using control samples
in data, as described below. This method greatly reduces the dependence of the
background prediction on the accurate modeling in simulation, the knowledge
of the inclusive \ttbar\ production cross-section, the measurement
of the integrated luminosity, and the accuracy of the object-selection
efficiency determination.

Additional backgrounds arise from processes involving one or more
$\PW$ and $\cPZ$ bosons, although these contributions are suppressed
by the \bhjet\ requirements. Finally, all event categories have backgrounds
from rare SM processes, such as $\ttbarZ$ and $\ttbarW$ production, whose cross sections have not been precisely measured~\cite{Chatrchyan:2013qca}. The prediction for
these contributions is derived from simulation, and a systematic uncertainty
of 50\% is assigned to account for the uncertainty in the NLO
calculations of their differential cross sections. The remainder of this section describes the
background predictions for each of the specific event categories.

\subsection{Backgrounds in event categories with a single lepton or two OS leptons}

For the single-lepton or two-OS-lepton event categories, the dominant
background is from $\ttbar$ events (85--95\% of the total). These
events can have three or more \bjets\ if the $\ttbar$ pair is accompanied by
additional jets that may be mistagged in the case of light-parton jets or
that may contain genuine \bjets\ from gluon decays to a $\bbbar$ pairs.
In the case of semileptonic \ttbar\ events, there are small
additional contributions from $\PW \to \cPqc \cPaqs$ decays,
with a charm-quark jet misidentified as a b jet, and from the
rare $\PW \to \cPqc \cPaqb$ decay mode.
In the case of dileptonic \ttbar\ events, $\tau$ leptons from the second
$\PW$ boson decay that are misidentified as \bjets\ also contribute.
Scale factors, defined as the ratio of the yield in data to the yield
in simulation, are used to normalize the background predictions from
simulation. For each SR, the corresponding scale factor is derived from
a control region enhanced in background $\ttbar$ events.
These control regions are defined by $50 \le \mt \le 100$\GeV for
the single-lepton selections and by either $\nbb=0$ or $\nbb=1$,
with either $\mbb\le 100\GeV$ or $\mbb\ge150\GeV$, for
the OS-dilepton case.
The contribution from non-$\ttbar$ events is evaluated from simulation
and subtracted from the data before deriving the normalization.
To reduce the contribution from a possible signal in these
control regions, the samples are restricted to events with low jet
multiplicity: for the three-\bhjet\ category, only events with exactly
five jets are used, and for the category with four \bjets,
only events with exactly four jets are used.
The dominant source of uncertainty for the background
prediction arises from the limited number of events in the control
samples (15--35\% on the total background).
The $\ttbar$ background prediction also depends on the ratio of events in
the signal and control regions, which is evaluated from
simulation and validated using $\ttbar$-dominated
control samples obtained by selecting events with fewer than three b
jets, as described below.

In the single-lepton channel, the modeling of the high-$\mt$ tail is
critical for the background estimation.
Genuine semileptonic \ttbar\ events have an endpoint at
$\mt \approx m_{\PW}$, with \MET\ resolution effects primarily
responsible for populating the $\mt > m_{\PW}$ tail.
The effect of \MET\ resolution on the $\mt$ tails is
investigated by selecting events with one or two \bjets\ and by varying the number
of additional jets. The comparison of simulation with data in the
$\mt$ tail region is used to extract scale factors and uncertainties for the semileptonic
$\ttbar$ prediction. The scale factors are in the range 1.1--1.2, depending on the
$\mt$ requirement, with corresponding uncertainties of 5--10\%.
The semileptonic background contributes
50--60\% of the total background in the single-lepton
SRs. Events from genuine dileptonic \ttbar\ events can also satisfy the
single-lepton event selection if the second lepton is not identified
or is not isolated and can give rise to large values of $\MET$ and
$\mt$ due to the presence of two neutrinos. This \ttdl\ contribution
constitutes $\sim$30--40\% of the total background and is derived from
simulation, with scale factors consistent with unity, as determined
from comparison of data with simulation in the dilepton control regions.

In the channels with two OS leptons, the most important issues for the
background prediction are related to the construction of $\bhjet$ pairs (see
Section~\ref{sec:1l2OSlSelection} for the full list of requirements).
Modeling of the emission of additional radiation leading to jets and gluon splitting to \bbbar pairs,
 and of effects such as $\tau$-lepton mistagging, \cPqc-quark-jet mistagging,
and \bhjet\ identification efficiency, can affect the $\mbb$ variable.
The modeling of these effects is validated using the statistically precise
single-lepton control sample with $50 \le \mt \le 100$\GeV, in which the \mbb\
distributions in data and simulation are compared as a function of the
\bhjet\ multiplicity. The ratio of the number of events satisfying the
$\nbb$ and $\mbb$ requirements that define the signal and control
regions is compared in data and simulation. This study is used to derive
scale factors, which are found to be consistent with unity,
and uncertainties corresponding to 20--30\% of the total background uncertainty.

\subsection{Backgrounds in the event category with two SS leptons}
For the SRs with two SS leptons, the background estimates
and uncertainties are derived following the procedures described in
Ref.~\cite{CMS-PAS-SUS-13-013}. There are three main categories of
backgrounds. Non-prompt leptons are produced
from heavy-flavor decays, misidentified hadrons, muons from the decay-in-flight of light
mesons, and electrons from unidentified photon conversions.
Charge misidentification arises mainly from electrons that undergo
severe bremsstrahlung in the tracker material, leading to a
misreconstruction of the charge sign. Finally, rare SM processes yielding two genuine SS
leptons (typically a \ttbar\ pair in association with an $\PH$, $\PW$, or $\cPZ$ boson)
can contribute significantly,
especially in SRs with tight selection requirements.
Backgrounds from non-prompt
leptons and rare SM processes dominate, each contributing
20--80\% of the total, while charge
misidentification contributes 1--5\%.

The background from non-prompt leptons is evaluated using the
event yield in a control sample in which the same analysis selections are
applied, except there is at least one lepton that passes a loose
lepton selection but fails the full set of tight identification and
isolation requirements. This observed yield is corrected by a
``tight-to-loose'' ratio, the probability that a loosely identified
non-prompt lepton also passes the full set of requirements. This
correction factor is in turn measured in a control sample of QCD
multijet events enriched in non-prompt leptons. The ratio is
obtained as a function of lepton $\pt$ and $\eta$.
The event kinematics and the various sources of non-prompt leptons are
different in the QCD multijet sample, where the tight-to-loose
ratio is measured, and the signal sample, where it is applied. This
gives rise to a systematic uncertainty in the non-prompt lepton
background estimate.
The charge misidentification background is obtained using a sample
of OS \Pe \Pe\ and \Pe \Pgm\ events that satisfy the full
kinematic selection weighted by the \pt- and $\eta$-dependent probability
of electron charge misassignment.
The systematic uncertainty of the total background prediction is
dominated by the uncertainties from rare SM processes and from events with a jet
misidentified as a prompt lepton (30--50\% of the total background).

\subsection{Backgrounds in the event category with at least three leptons}
For SRs with at least three leptons, there are two main
types of backgrounds.
In the off-$\cPZ$ SRs, the background with two
prompt leptons and an additional object misidentified as
a prompt lepton dominates, comprising 50--90\% of the total. In the
on-$\cPZ$ SRs, the dominant background is typically
from SM processes with at least three genuine prompt leptons,
corresponding to 60--100\% of the total.

The background sources with two prompt leptons from $\PW$ or $\cPZ$ boson decay
and a third object misidentified as a prompt lepton are
predominantly from $\ttbar$ production, although the $\zjets$ and
$\wwjets$ processes also contribute. The procedure to estimate this background
contribution follows closely that used for the analysis of events with
two SS leptons~\cite{CMS-PAS-SUS-13-013}.
The probability for a loosely identified lepton to satisfy the full set of selection
requirements is applied to a sample of $\ge$ 3 $\ell$ events,
in which the isolation requirement on one of the leptons is removed,
providing an estimate of the background contribution from
non-prompt leptons. A systematic
uncertainty of 30\% is derived for this background based on studies
of the method in simulation. This uncertainty accounts for the
difference in the $\pt$ spectrum of \bjets\ in the control
sample, where the probability is measured, compared to the spectrum in
the signal sample, where it is applied.
This systematic uncertainty
dominates the uncertainty in the background prediction in the SRs with
looser kinematic requirements.
SRs with tight kinematic requirements
also have a significant statistical uncertainty due to the size of the
sample used to derive this background estimate. These are the dominant
sources of uncertainty in the backgrounds in the off-$\cPZ$ signal
regions, corresponding to 20--90\% uncertainty on the total background.

The background contribution from events with two vector bosons that
produce three genuine prompt isolated leptons, mainly $\wzjets$ and
$\zzjets$ events, is estimated from simulation and is validated by comparing
data and simulation in control samples in which the full selection is
applied and the \bhjet\ requirement is inverted.
A control sample enhanced in the $\PW\cPZ$ background is obtained by
selecting events with three high-$\pt$ leptons. One pair of leptons is
required to form a $\cPZ\to\ell^+\ell^-$ candidate. The
third lepton is combined with the \MET vector, and this system is
required to form a $\PW$ boson candidate
($50<\ETmiss<100\GeV$ and $50<\MT<120\GeV$).
A second control sample, enhanced in the $\cPZ\cPZ$
background, is obtained by selecting events with four leptons and
$\ETmiss <$ 50\GeV. Two leptons are required to form a $\cPZ$
candidate. Scale factors are derived based on the comparison of data
and simulation in these control samples. The scale factors are found
to be unity and 0.9 for the $\PW\cPZ$ and $\cPZ\cPZ$ backgrounds,
respectively. The systematic uncertainty for the diboson background
is derived based on these comparisons, which are limited by the
statistical precision of the control samples. A 50\% uncertainty is
assigned to account for possible mismodeling of additional partons
required to satisfy the \bhjet\ requirement.

\section{Results}
\label{sec:results}

\begin{figure*}[tbh]
\centering
\includegraphics[width=0.7\textwidth]{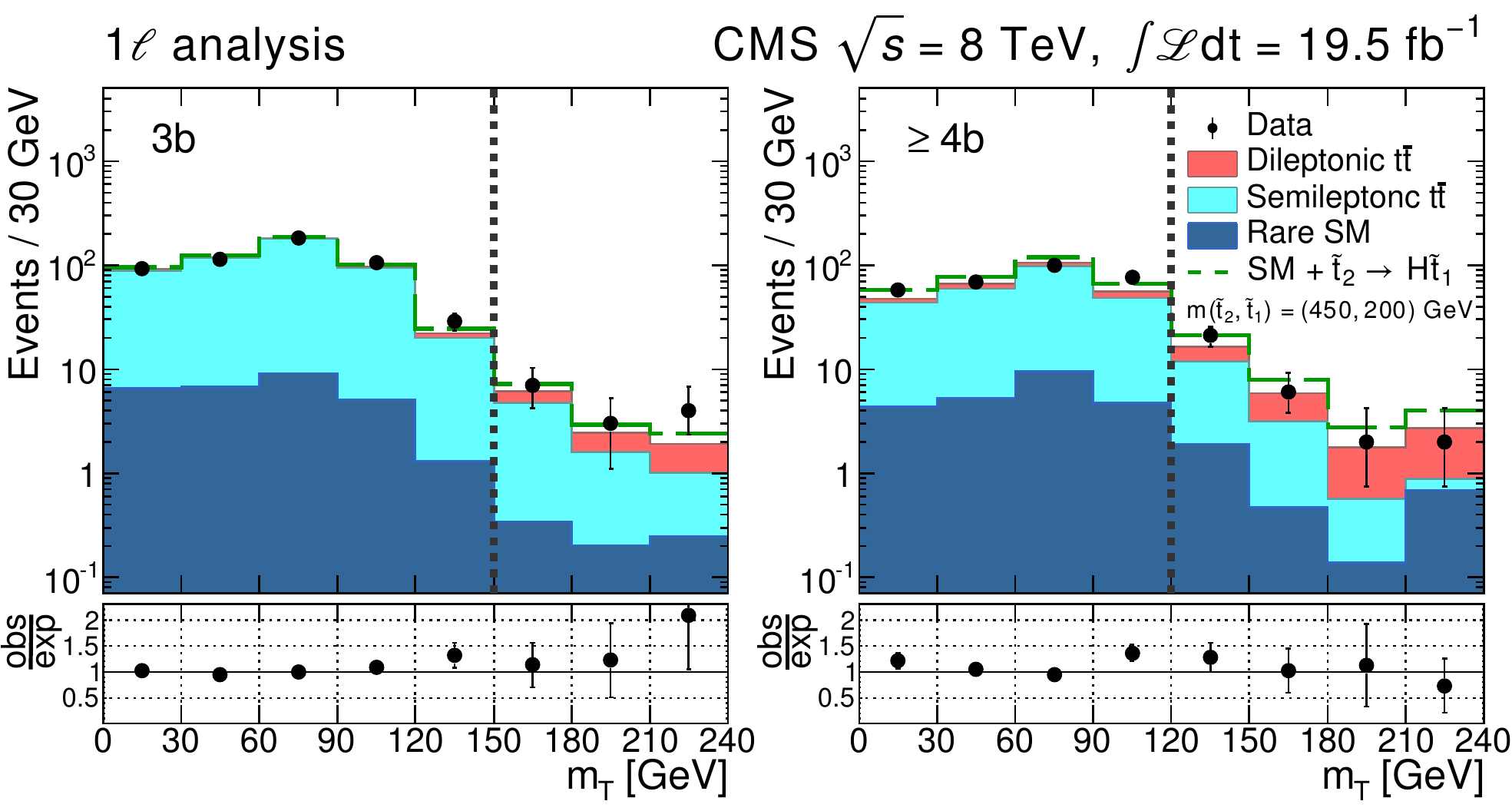}
\includegraphics[width=0.7\textwidth]{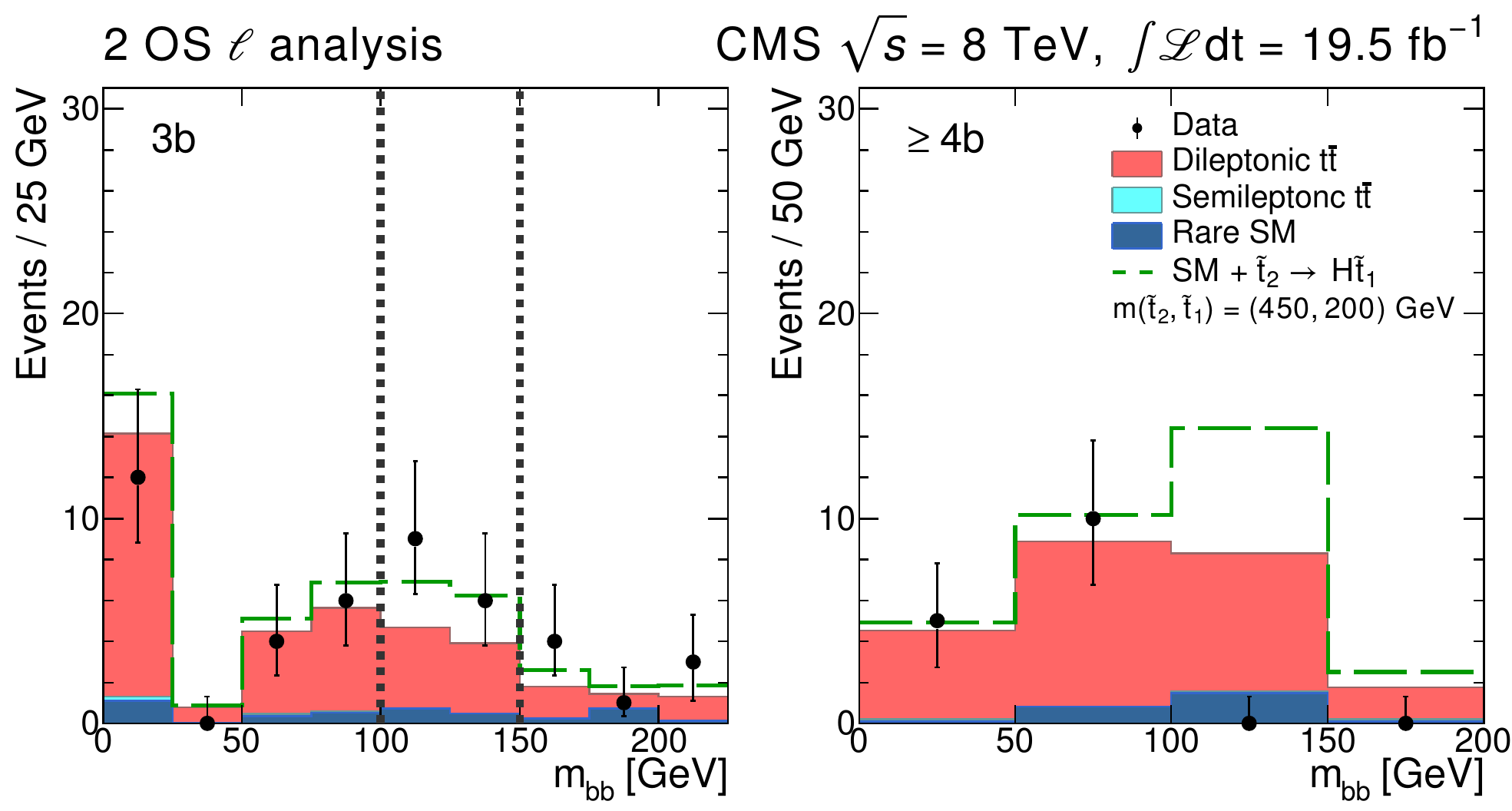}
\caption{
Comparison of the \mt\ distributions for events with one lepton (top row)
and \mbb\ distributions for events with two OS leptons (bottom row) in data
and MC simulation satisfying the $3\cPqb$ (left) and $\ge 4\cPqb$
(right) SR requirements.
The vertical dashed lines indicate the corresponding
signal region requirement.
The semileptonic $\ttbar$ and dileptonic $\ttbar$ components
represent simulated events characterized by the presence of one or two
$\PW$ bosons decaying to $\Pe$, $\mu$ or $\tau$.
The yields of the $\ttbar$ simulated samples are adjusted
so that the total SM prediction is normalized to the data in
the samples obtained by inverting the SR requirements.
The distribution for the model $\sttwo \to \PH \stone$ where
$m_{\sttwo}=450$\GeV and $m_{\stone}=200$\GeV is displayed on top of
the backgrounds.
The last bin contains the overflow events. The uncertainties
in the background predictions are derived for the total
yields in the signal regions and are listed in Table~\ref{tab:results_1l2l}.}
\label{fig:signalregions_1l2l}
\end{figure*}

\begin{figure*}[htb]
\centering
\includegraphics[width=0.90\textwidth]{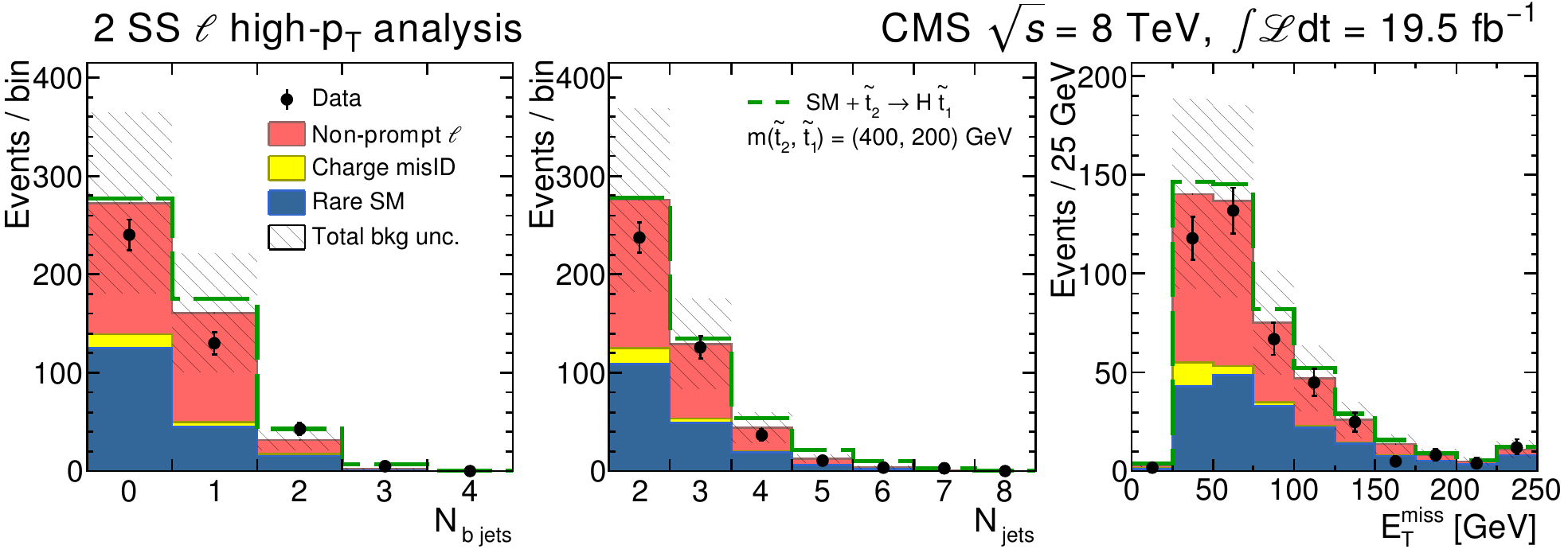}
\includegraphics[width=0.90\textwidth]{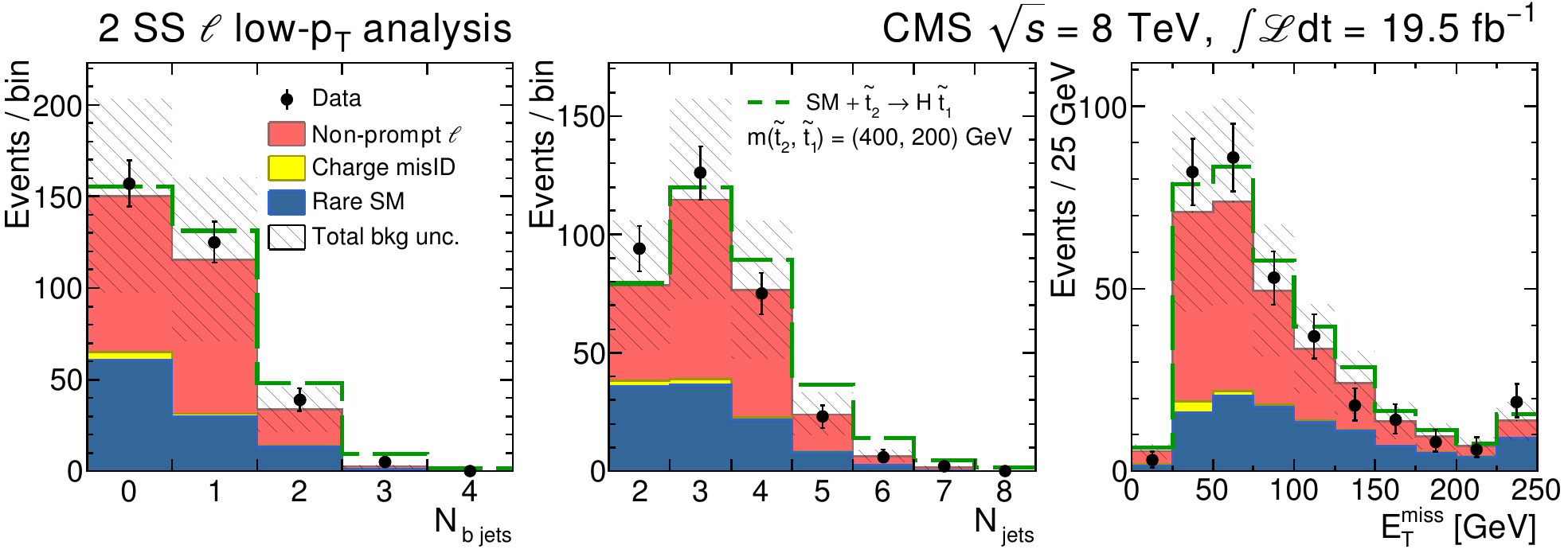}
\caption{Data and predicted SM background for the
  event sample with two SS leptons as a function of number of \bjets,
  number of jets, and $\ETmiss$ for events satisfying the
  high-$\pt$ (top row) or the low-$\pt$ (bottom row) selection. The shaded bands
  correspond to the total estimated uncertainty in the background
  prediction. The distribution for the model $\sttwo \to \PH \stone$ where
  $m_{\sttwo}=400$\GeV and $m_{\stone}=200$\GeV is displayed on top of
  the backgrounds. The last bin in the histograms includes
  overflow events.}
\label{fig:Baseline_SS}
\end{figure*}

\begin{figure*}[htb]
\centering
\includegraphics[width=0.90\textwidth]{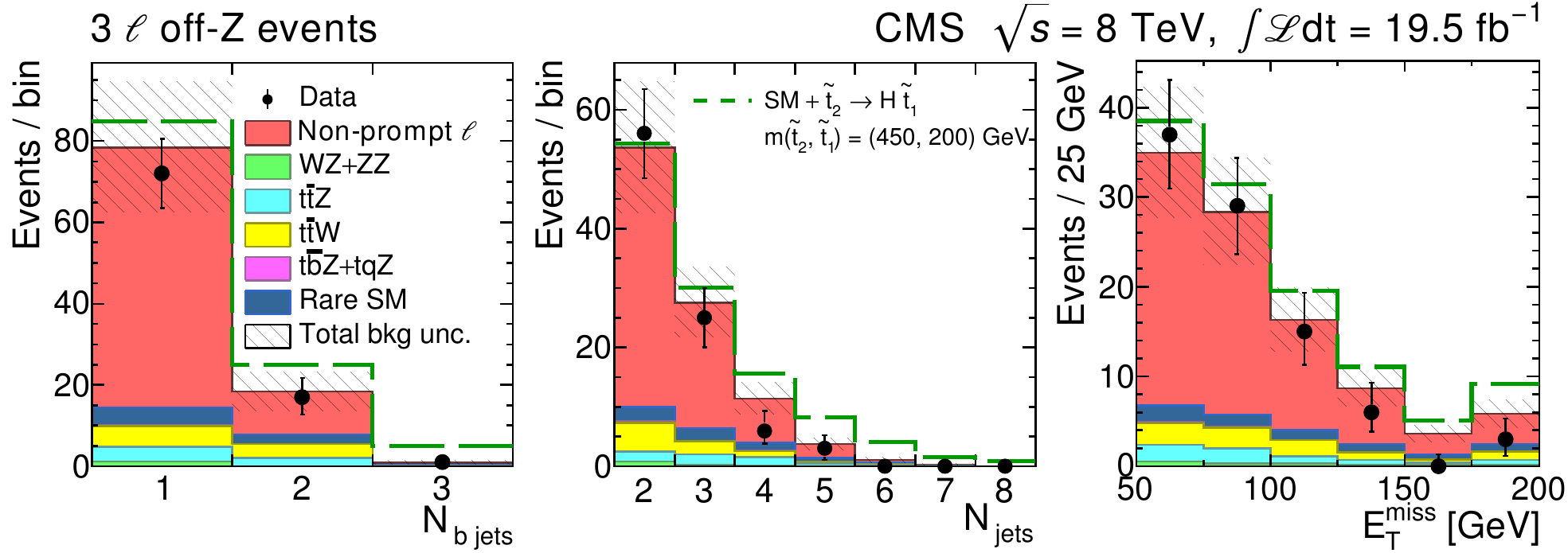}
\includegraphics[width=0.90\textwidth]{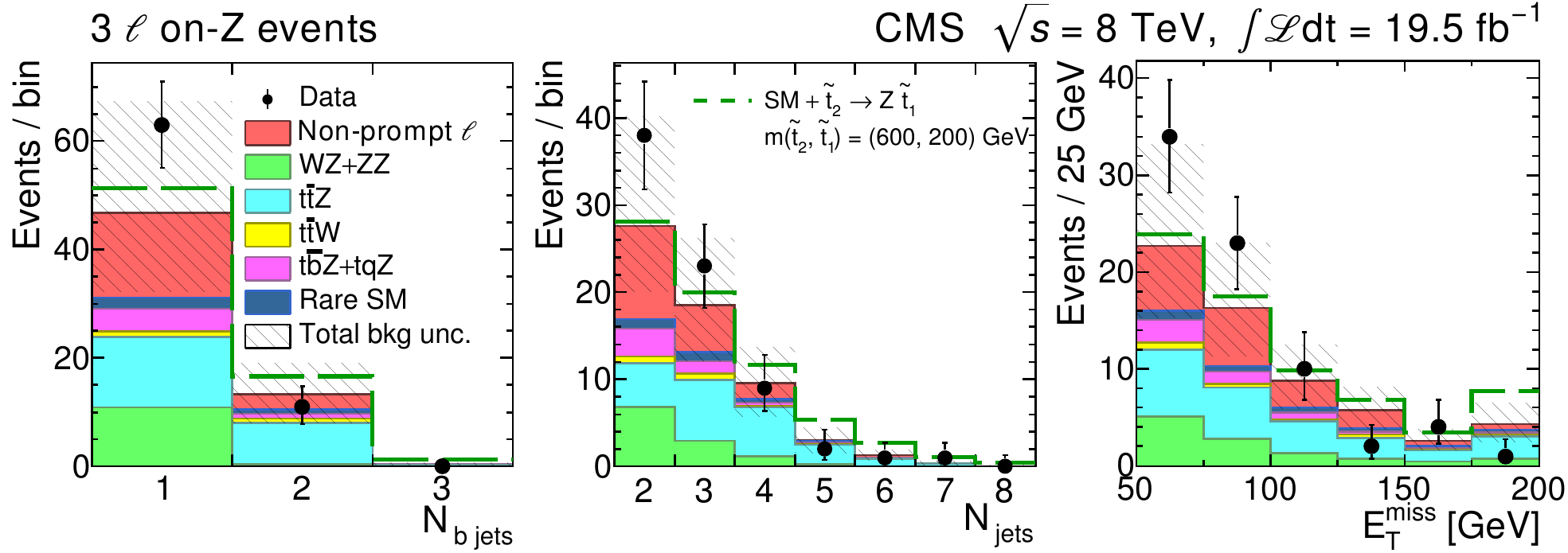}
\caption{Data and predicted SM background for the
  event sample with at least three leptons as a function of number of \bjets,
  number of jets, and $\ETmiss$ for events that
  do not contain (off-$\cPZ$), top row, or contain (on-$\cPZ$), bottom
  row, an OS same-flavor pair that is a $\cPZ$ boson
  candidate. The shaded bands correspond to the total estimated
  uncertainty in the background prediction. The distributions for the
  models $\sttwo \to \PH \stone$ and $\sttwo \to \cPZ
  \stone$ are displayed on top of the backgrounds in the top and
  bottom rows respectively. The top-squark masses are
  $m_{\stone}=200$\GeV and $m_{\sttwo}=(450, 600)$\GeV for the
  $(\PH,\cPZ)$ channel. The last bin in the histograms includes
  overflow events.}
\label{fig:Baseline}
\end{figure*}

The results of the search are shown in
Tables~\ref{tab:results_1l2l}-\ref{tab:results_3l}, and in
Figs.~\ref{fig:signalregions_1l2l}-\ref{fig:Baseline}, where the
background predictions are broken down into the various components.

\begin{table*}[bth]
\centering
\topcaption{Selection with one lepton or two OS leptons:
  background predictions and observed data yields. 
  The uncertainties in the total background predictions
  include both the statistical and systematic components. \label{tab:results_1l2l}}
\begin{tabular}{c c c |cc|cc}
\hline
\multirow{2}{*}{\nbjets} &
\multirow{2}{*}{\njets}&\multirow{2}{*}{\MET [\GeVns{}]}  &
\multicolumn{2}{c|}{$1\ell$ high $\MT$} &
\multicolumn{2}{c}{2 OS $\ell$  and $\cPqb\cPqb$ requirement} \\
& &  & Bkg. & Obs.  & Bkg. & Obs.  \tablespace \\
\hline
$=$3 & $\ge$5 & \multirow{2}{*}{$\ge$50} & $10.0 \pm 1.8$ & $14 $ & $8.4 \pm 2.7$ &$15 $ \\
$\ge$4 & $\ge$4 & & $27 \pm 6$ & $31 $ & $11 \pm 5\phantom{0}$ &$3 $ \\
\hline
\end{tabular}
\end{table*}

For the event selections with one lepton,
Fig.~\ref{fig:signalregions_1l2l} (top) 
shows a comparison of the $\mt$ distribution in
data and simulation.
The sample at low $\mt$ is enhanced in
semileptonic $\ttbar$ events and is used as a control sample
to derive the normalization for this background
contribution.
As shown in Fig.~\ref{fig:signalregions_1l2l} (top), the backgrounds in the
SR are mainly semileptonic and dileptonic \ttbar\ events.

For the SRs with two OS leptons,
Fig.~\ref{fig:signalregions_1l2l} (bottom)
shows a comparison of the $\mbb$ distribution in data and simulation.
The sample in the region outside the $\mbb$ signal window is used to
derive the normalization for the \ttdl\ background prediction for
events with three \bjets. In the case of events with at least four b
jets, multiple $\cPqb$$\cPqb$ pairs are possible. The control
region is not indicated in Fig.~\ref{fig:signalregions_1l2l} (bottom right)
since the $\mbb$ requirement is not applied when $\nbb\ge2$. 

The dominant background in the SRs is from \ttdl\ events.
The results for the SRs with one lepton or two OS
leptons are summarized in Table~\ref{tab:results_1l2l}. The predicted
and observed yields agree within 1.4 standard deviations of local
significance~\cite{2008NIMPA.595..480C}, given the statistical
uncertainty in the predicted yields.

\begin{table*}[htb]
\centering
\topcaption{SS dilepton event category: predicted total background and observed data yields as a
    function of the jet multiplicity, b-jet multiplicity, $\ETmiss$,
    and $\HT$ requirements, for the low-\pt and high-\pt regions.
    The uncertainties in the total background predictions
  contain the statistical and systematic components. \label{tab:results_ss}}
\resizebox{\textwidth}{!}{
\begin{tabular}{c c c |cc cc| cc cc}
 \multicolumn{3}{c}{Selection} & \multicolumn{4}{c}{low-$\pt$} & \multicolumn{4}{c}{high-$\pt$} \\
\hline
\multirow{2}{*}{\nbjets} & \multirow{2}{*}{\njets} & \multirow{2}{*}{\MET [\GeVns{}]}  & \multicolumn{2}{c}{$\HT\in[250,400]\GeV$} & \multicolumn{2}{c|}{$\HT\ge400\GeV$} & \multicolumn{2}{c}{$\HT\in[200,400]\GeV$} & \multicolumn{2}{c}{$\HT\ge400\GeV$} \tablespace \\
 &  &   & Bkg. & Obs. & Bkg. & Obs. & Bkg. & Obs. & Bkg. & Obs. \\
\hline\hline
	 \multirow{4}{*}{$=1$} 	 & \multirow{2}{*}{2--3} 	& 50--120 & 	 $29 \pm 12$ & 	 39 & 	 $5.6 \pm 2.0$ & 	 5 & 	 $31 \pm 12$ & 	 27 & 	 $3.4 \pm 1.2$ & 	 5 \\
	  	 &  	 & $\ge$120 & 	 $11 \pm 4\phantom{0}$ & 	 8 & 	 $4.9 \pm 1.8$ & 	 5 & 	 $9.0 \pm 3.2$ & 	 9 & 	 $3.5 \pm 1.3$ & 	 2 \\ \cline{2-11}
	  	 & \multirow{2}{*}{$\ge$4} 	 & 50--120 & 	 $15 \pm 6\phantom{0}$ & 	 15 & 	 $10 \pm 4\phantom{0}$ & 	 6 & 	 $9.2 \pm 3.4$ & 	 6 & 	 $5.4 \pm 2.0$ & 	 2 \\
	  	 & 	 & $\ge$120 & 	 $3.9 \pm 1.5$ & 	 3 & 	 $6.1 \pm 2.2$ & 	 10 & 	 $2.6 \pm 1.0$ & 	 3 & 	 $3.5 \pm 1.3$ & 	 6 \\
\hline
	 \multirow{4}{*}{$\ge$2} 	 & \multirow{2}{*}{2--3} 	 & 50--120 & 	 $6.6 \pm 2.4$ & 	 10 & 	 $1.3 \pm 0.5$ & 	 1 & 	 $6.0 \pm 2.1$ & 	 11 & 	 $0.78 \pm 0.34$ & 	 1 \\
	  	 & 	 & $\ge$120 & 	 $2.4 \pm 0.9$ & 	 1 & 	 $1.2 \pm 0.5$ & 	 2 & 	 $2.4 \pm 0.9$ & 	 3 & 	 $0.8 \pm 0.4$ & 	 1 \\ \cline{2-11}
	  	 & \multirow{2}{*}{$\ge$4}  	 & 50--120 & 	 $6.5 \pm 2.5$ & 	 5 & 	 $4.0 \pm 1.5$ & 	 11 & 	 $3.4 \pm 1.3$ & 	 2 & 	 $2.3 \pm 1.0$ & 	 7 \\
	  	 & 	 & $\ge$120 & 	 $1.8 \pm 0.7$ & 	 0 & 	 $3.1 \pm 1.2$ & 	 3 & 	 $1.1 \pm 0.5$ & 	 0 & 	 $2.0 \pm 0.8$ & 	 2 \\
\hline
\end{tabular}
}
\end{table*}

Figure~\ref{fig:Baseline_SS} shows a comparison of data and the
predicted backgrounds for events with two SS leptons satisfying
a more inclusive selection, which is enhanced in SM processes: at least two
jets, moderate $\HT$ ($>$250\GeV in the low-$\pt$ analysis and
$>$80\GeV in the high-$\pt$ analysis), and moderate $\ETmiss$
($>$30\GeV for events with $\HT<500\GeV$; otherwise, there is
no $\ETmiss$ requirement). This sample serves to validate the
methods used to predict the backgrounds in the SRs, which are
defined by applying requirements on the
selection observables shown: the jet multiplicity, the \bhjet\
multiplicity, and $\MET$, as well as $\HT$.
The amount of background varies strongly among the signal regions;
some of them including tens of background events while others have
essentially none.
The relative contribution from rare SM processes
increases as the requirements are tightened.
As shown in Table~\ref{tab:results_ss}, the SM background predictions
and observations in the SRs are in agreement for both the
high-$\pt$ and low-$\pt$ selections.

\begin{table*}[tbh]
\centering
  \topcaption{Predicted total background and observed data yields as a
    function of the jet multiplicity, b-jet multiplicity, $\ETmiss$,
    and $\HT$ requirements, for events with at least three leptons,
    with (on-$\cPZ$) and without (off-$\cPZ$) a $\cPZ$ boson candidate
    present. The uncertainties in the total background predictions
    include both the statistical and systematic components.
\label{tab:results_3l}}
\resizebox{\textwidth}{!}{
\begin{tabular}{c c c |cc cc |cc cc}
 \multicolumn{3}{c}{Selection} & \multicolumn{4}{c}{off-$\cPZ$} & \multicolumn{4}{c}{on-$\cPZ$} \\
\hline
\multirow{2}{*}{\nbjets} & \multirow{2}{*}{\njets} & \multirow{2}{*}{\MET [\GeVns{}]}  & \multicolumn{2}{c}{$\HT\in[60,200]\GeV$} & \multicolumn{2}{c|}{$\HT\ge200\GeV$} & \multicolumn{2}{c}{$\HT\in[60,200]\GeV$} & \multicolumn{2}{c}{$\HT\ge200\GeV$}  \tablespace \\
 &  &   & Bkg. & Obs. & Bkg. & Obs. & Bkg. & Obs. & Bkg. & Obs.  \tablespace \\
\hline\hline
	 \multirow{6}{*}{$=1$} 	 & \multirow{3}{*}{2--3} 	 & 50--100 & 	 $34 \pm 7\phantom{0}$ & 	 36 & 	 $11.2 \pm 2.5\phantom{0}$ & 	 9 & 	 $16 \pm 5\phantom{0}$ & 	 30 & 	 $10 \pm 4\phantom{0}$ & 	 13 \\
	  	 &  	 & 100--200 & 	 $12.2 \pm 2.7\phantom{0}$ & 	 13 & 	 $9.1 \pm 2.1$ & 	 6 & 	 $5.3 \pm 1.8$ & 	 6 & 	 $5.9 \pm 2.1$ & 	 3 \\
	  	 &  	 & $\ge$200 & 	 $0.33 \pm 0.22$ & 	 0 & 	 $1.2 \pm 0.5$ & 	 0 & 	 $0.37 \pm 0.23$ & 	 0 & 	 $0.9 \pm 0.4$ & 	 0 \\\cline{2-11}
	  	 &  \multirow{3}{*}{$\ge$4} 	 & 50--100 & 	 $0.9 \pm 0.4$ & 	 2 & 	 $5.4 \pm 1.3$ & 	 3 & 	 $0.11 \pm 0.13$ & 	 1 & 	 $5.0 \pm 2.0$ & 	 4 \\
	  	 &  	 & 100--200 & 	 $0.10 \pm 0.12$ & 	 0 & 	 $3.6 \pm 1.0$ & 	 3 & 	 $0.08 \pm 0.12$ & 	 0 & 	 $3.0 \pm 1.3$ & 	 5 \\
	  	 &  	 & $\ge$200 & 	 $0.0 \pm 0.1$ & 	 0 & 	 $0.76 \pm 0.35$ & 	 0 & 	 $0.02 \pm 0.10$ & 	 0 & 	 $0.56 \pm 0.32$ & 	 1 \\
\hline
	 \multirow{6}{*}{$=2$} 	 & \multirow{3}{*}{2--3} 	 & 50--100 & 	 $4.9 \pm 1.2$ & 	 7 & 	 $3.9 \pm 1.2$ & 	 7 & 	 $2.4 \pm 0.9$ & 	 5 & 	 $2.5 \pm 1.1$ & 	 2 \\
	  	 &  	 & 100--200 & 	 $2.3 \pm 0.7$ & 	 1 & 	 $1.9 \pm 0.7$ & 	 0 & 	 $1.3 \pm 0.5$ & 	 1 & 	 $1.4 \pm 0.6$ & 	 1 \\
	  	 &  	 & $\ge$200 & 	 $0.22 \pm 0.21$ & 	 1 & 	 $0.14 \pm 0.14$ & 	 0 & 	 $0.12 \pm 0.13$ & 	 0 & 	 $0.43 \pm 0.26$ & 	 0 \\\cline{2-11}
	  	 &  \multirow{3}{*}{$\ge$4} 	 & 50--100 & 	 $0.03 \pm 0.11$ & 	 0 & 	 $2.8 \pm 0.9$ & 	 1 & 	 $0.20 \pm 0.17$ & 	 1 & 	 $2.9 \pm 1.3$ & 	 1 \\
	  	 &  	 & 100--200 & 	 $0.05 \pm 0.11$ & 	 0 & 	 $1.7 \pm 0.6$ & 	 0 & 	 $0.10 \pm 0.13$ & 	 0 & 	 $1.7 \pm 0.8$ & 	 0 \\
	  	 &  	 & $\ge$200 & 	 $0.0 \pm 0.1$ & 	 0 & 	 $0.38 \pm 0.21$ & 	 0 & 	 $0.0 \pm 0.1$ & 	 0 & 	 $0.29 \pm 0.19$ & 	 0 \\
\hline
	 \multirow{3}{*}{$\ge$3} 	 & \multirow{3}{*}{$\ge$3} 	 & 50--100 & 	 $0.0 \pm 0.1$ & 	 0 & 	 $0.56 \pm 0.27$ & 	 1 & 	 $0.0 \pm 0.1$ & 	 0 & 	 $0.18 \pm 0.15$ & 	 0 \\
	  	 &  	 & 100--200 & 	 $0.02 \pm 0.11$ & 	 0 & 	 $0.18 \pm 0.14$ & 	 0 & 	 $0.0 \pm 0.1$ & 	 0 & 	 $0.25 \pm 0.17$ & 	 0 \\
	  	 &  	 & $\ge$200 & 	 $0.0 \pm 0.1$ & 	 0 & 	 $0.2 \pm 0.2$ & 	 0 & 	 $0.0 \pm 0.1$ & 	 0 & 	 $0.02 \pm 0.10$ & 	 0 \\
\hline
\end{tabular}
}
\end{table*}

Finally, for the event sample with at least three leptons,
Fig.~\ref{fig:Baseline} shows a comparison of data and the
predicted backgrounds for the jet and \bhjet\ multiplicities and for the $\MET$ distribution.
The dominant background is from processes with two prompt leptons and
additional non-prompt leptons, mainly due to $\ttbar$ events, although in
the case of the on-$\cPZ$ selection, background sources with $\cPZ$
bosons also contribute significantly.
The results of the search, summarized in Table~\ref{tab:results_3l}, demonstrate
agreement between background predictions and observations for all the
SRs considered.

In summary, the data yields are found to be consistent with the background predictions across all
event categories and SRs. Of the 96 SRs, the largest
discrepancy corresponds to a 1.6 standard deviation excess of local significance
(30 events compared to $16 \pm 5$ expected, see Table~\ref{tab:results_3l}), computed following the recommendations of Ref.~\cite{2008NIMPA.595..480C}.
Thus, no indication of top-squark pair production is observed.

\section{Interpretation}
\label{sec:interpretation}

The results are used to set upper limits on the cross section
times branching fraction for pair production of
$\sttwo$ squarks for the decay modes shown in Fig.~\ref{fig:T6ttXXdiagram}.
The upper limits are calculated
at a 95\% confidence level (CL) using the LHC-style CL$_\mathrm{S}$
method~\cite{Junk:1999kv,Read:2002hq,LHC-HCG}.
The exclusion curves on particle masses at 95\% CL are evaluated from a comparison of
the cross section upper limits and the theoretical signal cross section
predictions.
As explained below, the results from the various SRs are combined in
the limit-setting procedure in order to improve the sensitivity of the search.

The limit calculation on the cross section times branching fraction
depends on the signal selection efficiency and the background estimates.
The SRs with at least three leptons have the highest
expected sensitivity because of the small level of SM background.
For SRs with at least three leptons, the off-$\cPZ$ SRs with $\HT > 200\GeV$ are used for the $\ttbar\PH\PH$ interpretation,
while both the off-$\cPZ$ and on-$\cPZ$ SRs with $\HT > 200\GeV$ are used for the $\ttbar\cPZ\cPZ$ interpretation.
The total signal acceptance for all SRs with at least three
leptons varies from around  0.4--0.5\% for the $\ttbar\PH\PH$
signal, to  1.2--1.5\% for the $\ttbar\cPZ\cPZ$ signal. The acceptance
for the most sensitive SR alone is
around $\sim$0.1\% for $\ttbar\PH\PH$ and approximately
three times larger for $\ttbar\cPZ\cPZ$. This difference in acceptance
is due to the larger leptonic branching fraction for $\cPZ$ boson
decays compared to $\PH$ boson decays.
The SRs with lower lepton multiplicities also have
sensitivity to the $\ttbar\PH\PH$ signal. All SRs
of the high-$\pt$ SS dilepton analysis are used in the limit
setting. While only the high-$\pt$ results are used in the
interpretation presented in this letter, the low-$\pt$ experimental results are
included in Table~\ref{tab:results_ss} for potential use in future interpretations.
In SRs with two SS leptons, the overall acceptance for
$\ttbar\PH\PH$ events is 0.3--0.5\%, where the most sensitive signal
regions contribute $\sim$0.15\%. In the case of SRs with
one lepton or two OS leptons, the acceptance for $\ttbar\PH\PH$ events is
approximately 0.2--0.4\%. The acceptances for the single-lepton and
dilepton final states are slightly lower for the $\ttbar\cPZ\cPZ$
signal.
Because of the large branching fraction for the $\PH \to
\bbbar$ decay mode, SRs with higher \bhjet\ multiplicity requirements
dominate the expected sensitivity for scenarios with $\PH$ bosons.
SRs with low \bhjet\ multiplicities are most sensitive for scenarios
with $\cPZ$ bosons.

The systematic uncertainties, listed in Table~\ref{tab:sigsyst}, are evaluated for the signal
selection efficiency in every SR and for every signal point separately.
The total uncertainty in the signal selection efficiency is in the 9--30\% range.
The dominant source of uncertainty depends on the SR and
decay mode considered. An important source of uncertainty arises from the estimation of
the trigger and lepton identification efficiencies, which are derived
using $\cPZ\to\ell^+\ell^-$ samples and contribute 6--13\%.
The uncertainty due to the knowledge of the energy scale of hadronic jets increases with
tighter kinematic requirements and corresponds to an uncertainty of
1--15\%. The uncertainty due to the knowledge of the $\bjet$
identification performance depends on the event properties, such as the jet flavor
and $\pt$ value, and gives rise to an uncertainty of 2--20\%.
For smaller differences between the $m_{\sttwo}$ and $m_{\stone}$ mass values,
uncertainties in the modeling of initial-state radiation (ISR) become important.
The uncertainty related to the PDFs on the acceptance is determined using the PDF4LHC
recommendations~\cite{Botje:2011sn} and contributes 2--5\%.
The corresponding uncertainty in the signal selection efficiency is of
3--15\%, increasing for smaller $m_{\sttwo}$--$m_{\stone}$ mass differences.
The systematic uncertainties, including their correlations, are treated
consistently in the different analyses. The correlations between the different
analyses have a small impact on the combined result.

\begin{table*}[tbhp]
\centering
\topcaption{Relative systematic uncertainties (in percent) in the signal
  yields for the different event
  selections: one lepton  (1 $\ell$), two OS leptons (2 OS $\ell$),
  two SS leptons (2 SS $\ell$), and at least three leptons
  (${\ge}$ 3 $\ell$). The range indicates the variation in the systematic
  uncertainty for the different decay channels and SRs considered.
\label{tab:sigsyst}
}
\begin{tabular}{ l  | c  c  c  c }  \hline
Source  &  1 $\ell$ [\%] & 2 OS $\ell$ [\%] & 2 SS $\ell$ [\%] &
${\ge}$ 3 $\ell$ [\%]  \\ \hline
Luminosity~\cite{LUMIPAS} &  \multicolumn{4}{c}{2.6}   \\
Pileup modeling &  \multicolumn{4}{c}{$<5$}   \\\cline{2-5}
Trigger  efficiency & 3 & 6 & 6 & 5  \\
Lepton identification and isolation efficiency & 5 & 10 & 10 & 12 \\
Jet energy scale modeling & 1--3 & 1--3 & 1--10 & 5--15\\
\bhjet\ identification~\cite{ref:btag} & 3--5 & 3--5 & 2--10 & 5--20 \\
ISR modeling~\cite{SUS-13-011} & 3--5 & 3--5 & 3--15 & 3--15 \\
PDFs & 5 & 5 & 2 & 4 \\\hline
Total  & 9--11 & 14--15 & 14--23 & 15--30  \\ \hline
\end{tabular}
\end{table*}

\begin{figure*}[htbp]
\centering
\includegraphics[width=0.50\textwidth]{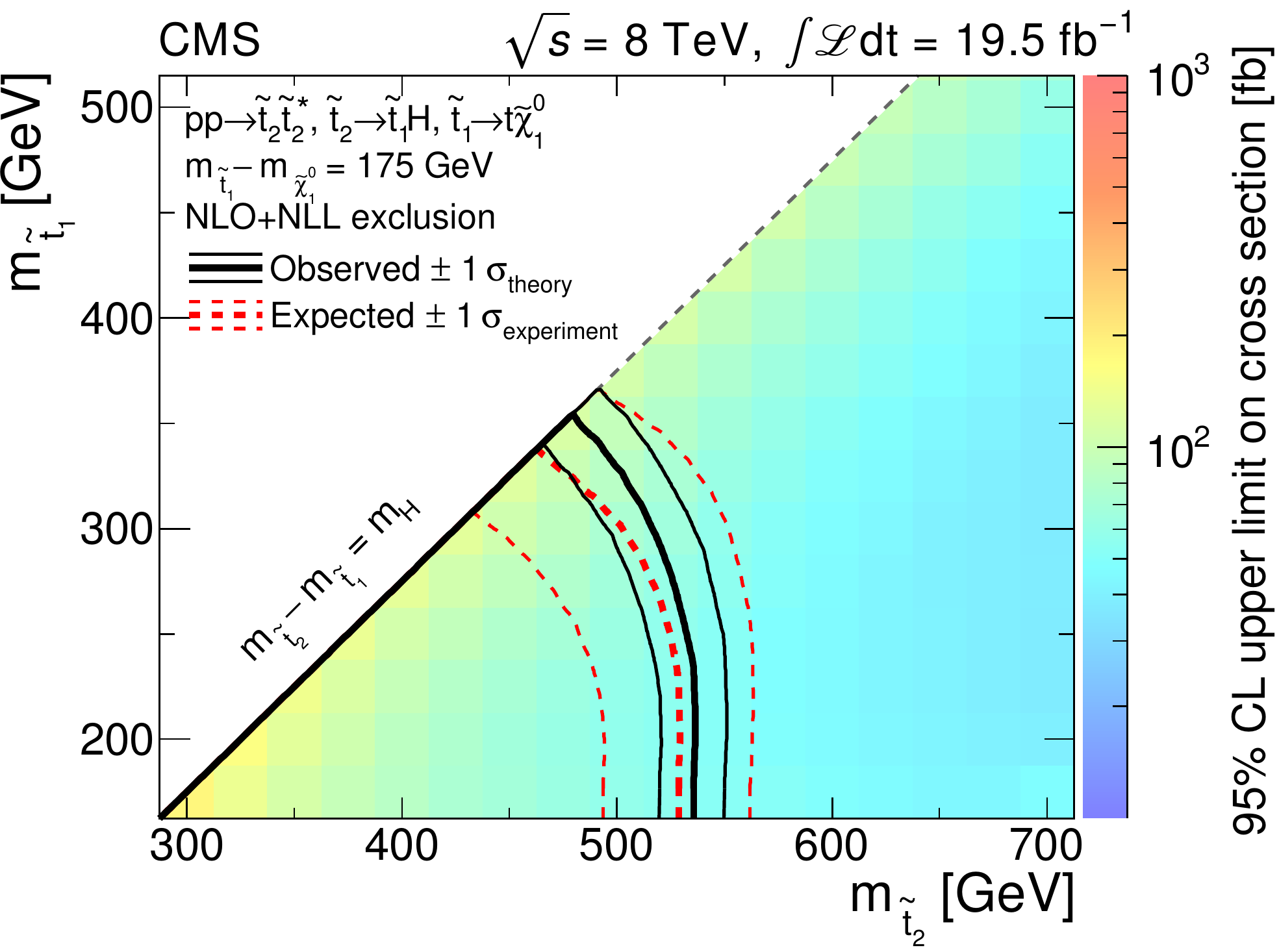}
\includegraphics[width=0.44\textwidth]{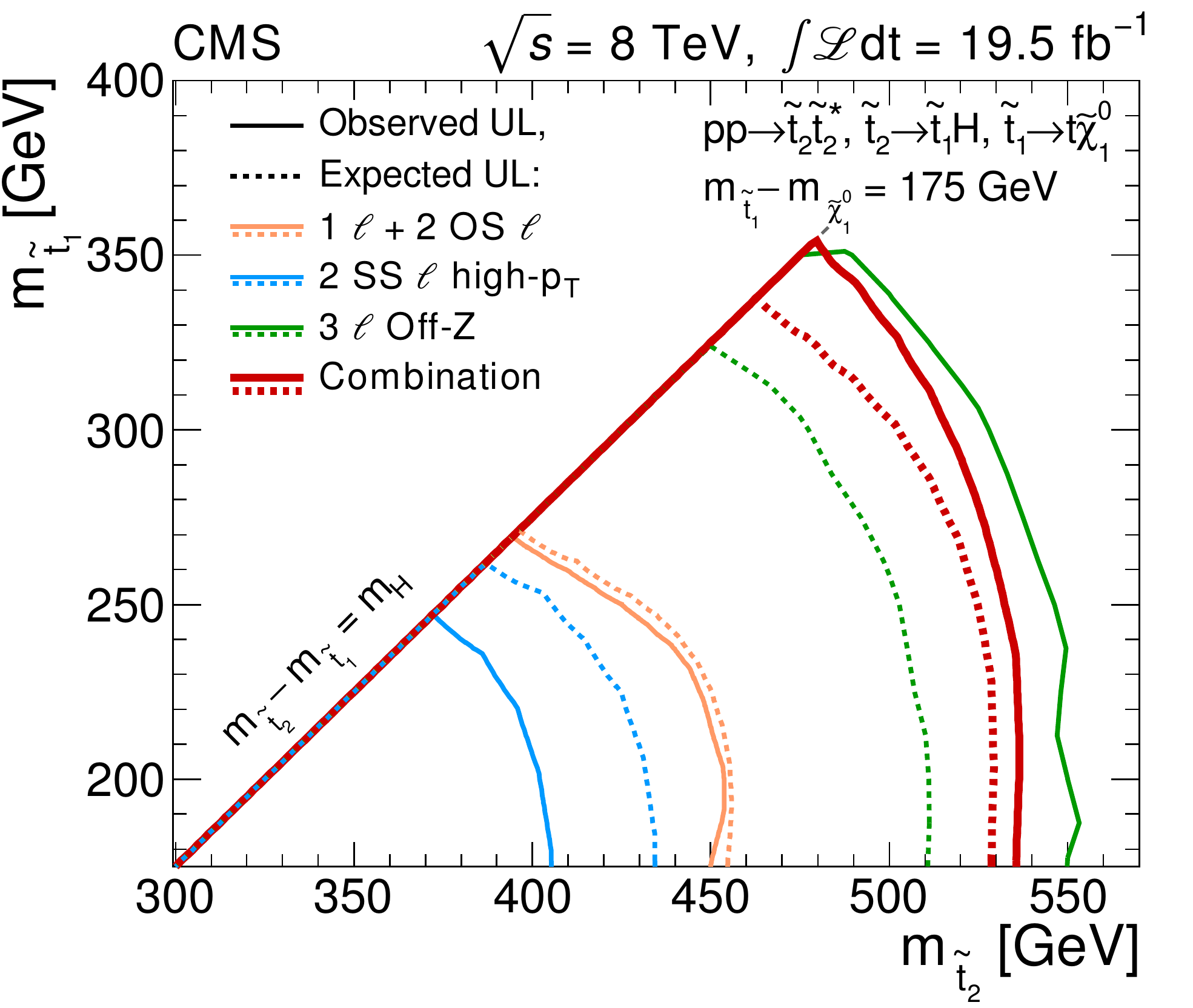}
\includegraphics[width=0.50\textwidth]{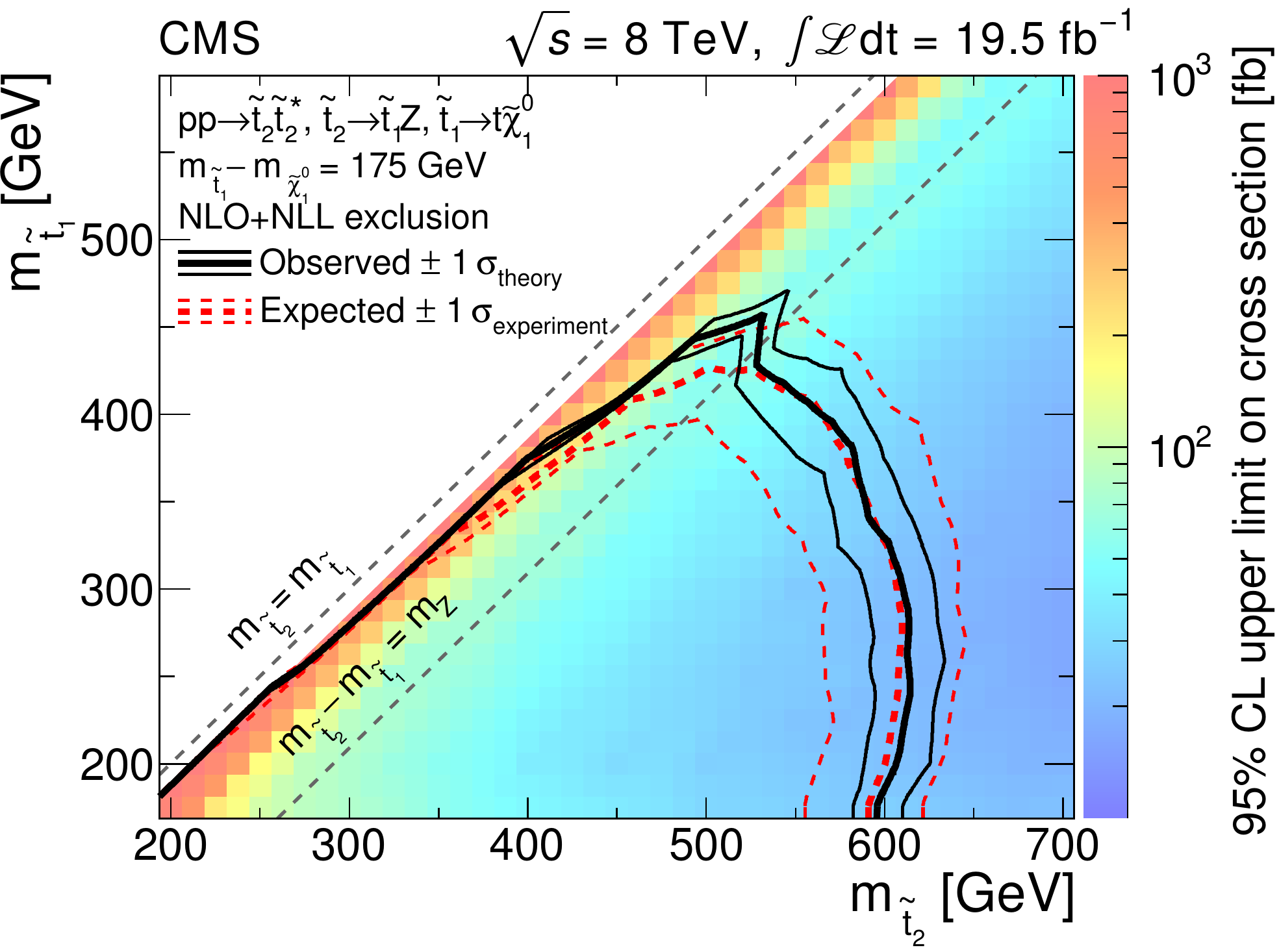}
\includegraphics[width=0.44\textwidth]{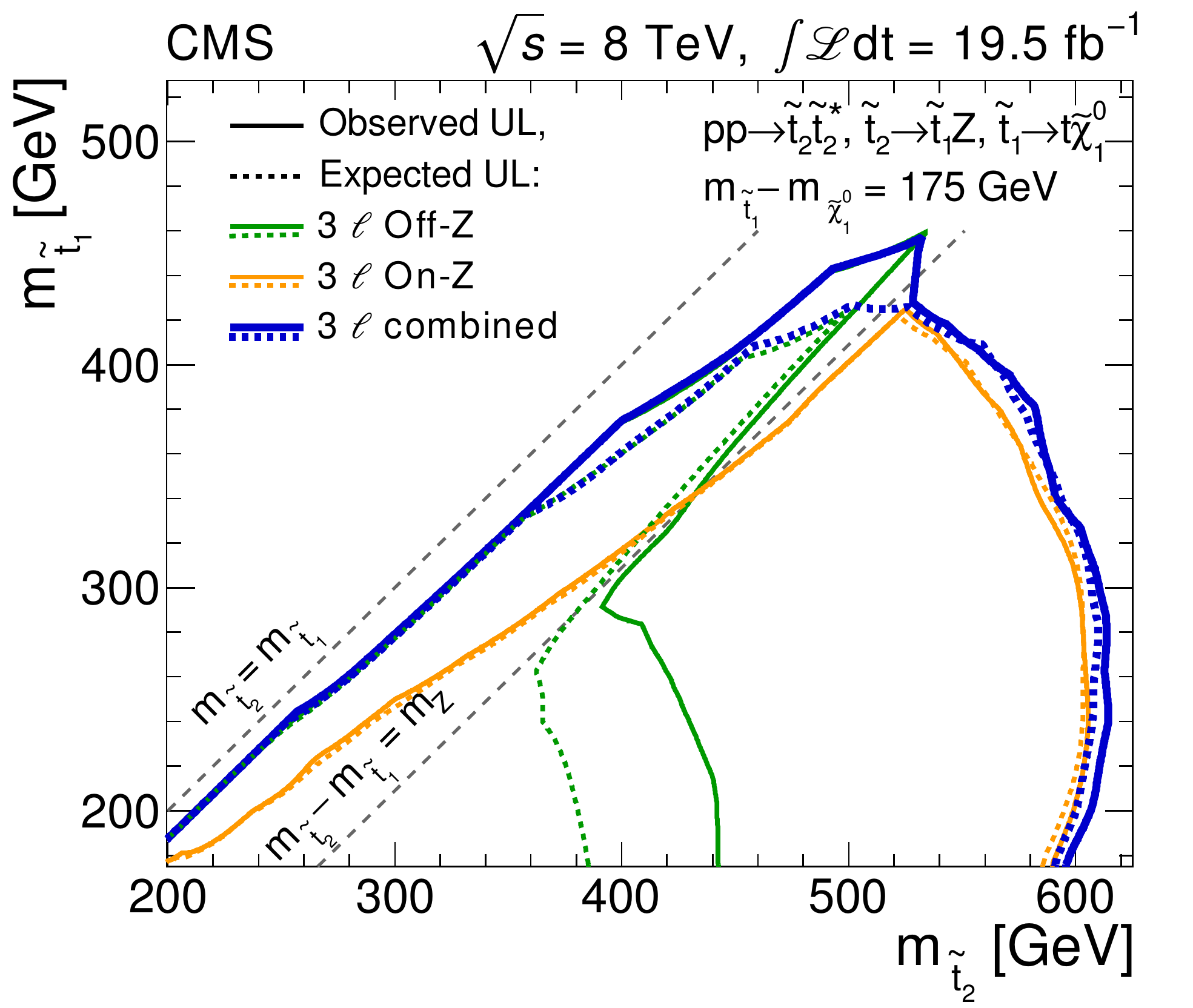}
\caption{
  Interpretation of the results in SUSY simplified model
  parameter space, $m_{\stone}$ \vs $m_{\sttwo}$, with the
  neutralino mass constrained by the relation
  $m_{\stone} - m_{\PSGczDo}  = 175\GeV$.
  The shaded maps (plots on the left) show the upper
  limit (95\% CL) on the cross section times branching fraction
  at each point in the $m_{\stone}$ \vs~$m_{\sttwo}$ plane for the
  process $\Pp\Pp \to \sttwo\sttwo^{*}$,
  with $\sttwo\to \PH \stone$, $\stone\to
  \cPqt\PSGczDo$ (upper plots) and $\PSQtDt\to \cPZ\PSQtDo$,
  $\stone\to \cPqt\PSGczDo$ (lower plots). In these plots,
  the results from all channels are combined. The excluded region
  in the $m_{\stone}$ \vs $m_{\sttwo}$ parameter space is obtained
  by comparing the cross section times branching fraction upper
  limit at each model point with the corresponding NLO+NLL
  cross section for the process, assuming that (a)~$\mathcal{B}(\sttwo\to \PH \stone)=100\%$ or (b)~that $\mathcal{B}(\sttwo\to \cPZ \stone)=100\%$.
  The solid (dashed) curves define the boundary of the observed
  (expected) excluded region. The $\pm$1 standard deviation
  ($\sigma$) bands are indicated by the finer contours. The figures
  on the right show the observed (expected) exclusion contours,
  which are indicated by the solid (dashed) curves for the
  contributing channels. As indicated in the legends of the
  right-hand figures, the thinner curves show the results from
  each of the contributing channels, while the thicker curve shows
  their combination. The four event categories for the
  $\sttwo\to \PH \stone$ study are shown in the upper plots, while the
  on-$\cPZ$ and off-$\cPZ$ categories for events with at least three
  leptons are shown in the lower plots.
  \label{fig:interp_HH_and_ZZ}}
\end{figure*}

Figure~\ref{fig:interp_HH_and_ZZ} (left) shows the 95\% CL upper limits
on the cross section times branching fraction in the $m_{\stone}$
versus $m_{\sttwo}$ plane for the (a)~$\sttwo\to \PH \stone$
and (b)~$\sttwo\to \cPZ \stone$ decay modes.
The contour bounds the excluded region
in the plane assuming the NLO+NLL cross section calculation in the
decoupling limit for all the SUSY sparticles not included in the
model. The results are presented assuming a branching fraction of 100\% to each decay mode.
The 95\% CL expected (thick dashed) and observed (solid black)
limits are obtained including all uncertainties with the exception of
the theoretical uncertainty in the signal production cross section.
The expected limit is defined as the median of the upper-limit
distribution obtained using pseudo-experiments and the likelihood model considered.
The bands around the expected limit correspond to the impact of
experimental uncertainties, and the bands around the
observed limit indicate the change for a $\pm$1 standard deviation ($\sigma$)
variation in the theoretical cross section (mainly due to
uncertainties in the renormalization/factorization scales and in the
knowledge of the PDFs).
In the $\sttwo\to \PH \stone$ decay mode,
taking a $-1\sigma$ theory lower bound on signal cross sections, a
$\sttwo$ squark with
$m_{\sttwo} \lesssim 525\GeV$ is excluded at a 95\% CL for $\stone$
squarks with $m_{\stone} \lesssim 300\GeV$.
Similarly, in the $\sttwo\to \cPZ \stone$ decay mode, a $\sttwo$ squark with
$m_{\sttwo} \lesssim 575\GeV$ is excluded at 95\% CL for $\stone$ squark with
$m_{\stone} \lesssim 400\GeV$.

For the pure $\sttwo\to \PH \stone$ decay (Fig.~\ref{fig:interp_HH_and_ZZ} upper right),
the SRs with at least three leptons, no $\cPZ\to \ell^+\ell^-$ candidates,
and large \bhjet\ multiplicities are the most sensitive.
Nevertheless, the SRs with lower lepton multiplicities
(one lepton or two leptons) have significant expected
sensitivity in the $\sttwo\to \PH \stone$ decay mode.
Including the final states with lower lepton
multiplicities in the combination lowers the cross section upper limit
results by 15--20\% compared to the
three-lepton results alone. Therefore, all lepton multiplicity categories are used
in the interpretation of the $\ttbar\PH\PH$ signal.

In the case of the signals with $\cPZ$ bosons
(Fig.~\ref{fig:interp_HH_and_ZZ} lower right), the SRs with
at least three leptons completely dominate the expected sensitivity.
The different SRs with at least three leptons provide
sensitivity to different types of signals.
In particular, off-$\cPZ$ SRs are sensitive to the region
of parameter space in which the $\cPZ$ bosons are off-shell,
$m_{\sttwo} - m_{\stone} < m_{\cPZ}$, while the on-$\cPZ$ regions
provide sensitivity to signals with larger mass
differences.
Only the SRs with at least three leptons are used in the interpretation of the $\ttbar\cPZ\cPZ$ signal.

\begin{figure}[hbtp]
\centering
\includegraphics[width=\cmsFigWidth]{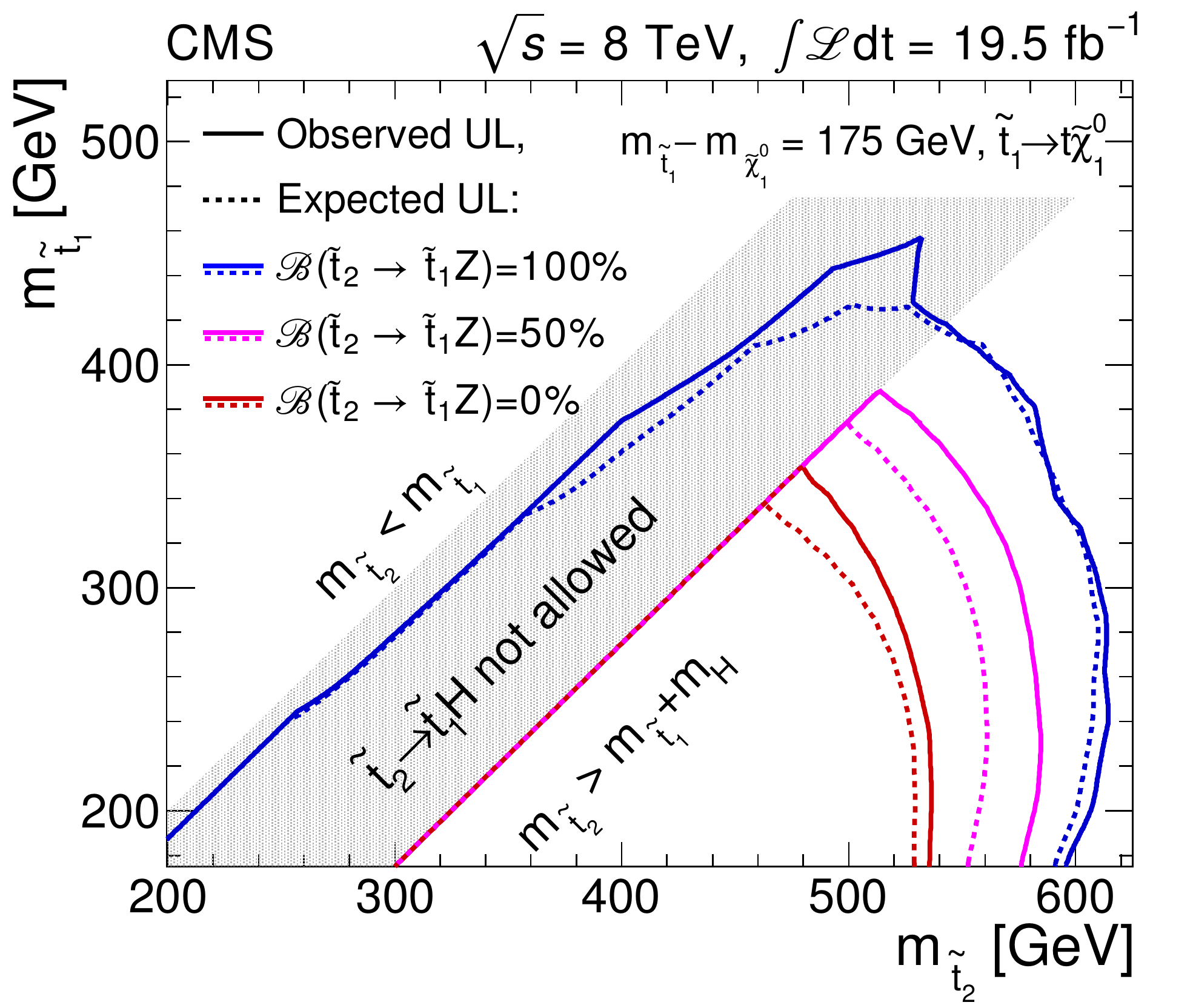}
\caption{Upper limits on the cross section for $\sttwo$ pair
  production for different branching fractions of $\sttwo\to \PH \stone$
  and $\sttwo\to \cPZ \stone$, assuming that
  $\mathcal{B}(\sttwo\to \PH \stone) +
  \mathcal{B}(\sttwo\to \cPZ \stone) = 100\%$.
  The $\stone$ squark is assumed to always decay to a top quark and a neutralino $\PSGczDo$ with
  $m_{\stone} - m_{\PSGczDo} = m_{\cPqt}$. The decay $\sttwo\to \PH \stone$ is
  only considered when the $\PH$ boson production is kinematically
  allowed, $m_{\sttwo}-m_{\stone}>m_{\PH}$. \label{fig:interp_HH_and_ZZ_BF}}
\end{figure}

Mixed-decay scenarios, with non-zero branching fractions for the $\cPZ$ and
$\PH$ decay modes, are also considered, assuming these to be the only
decay modes possible.
Figure~\ref{fig:interp_HH_and_ZZ_BF} shows the corresponding limits as
a function of the relative branching fraction of the $\cPZ$ and $\PH$
decay modes. The scenario with the least expected sensitivity is where the $\PH$
boson decay mode dominates, while the best expected sensitivity is achieved when
the $\cPZ$ boson decay mode dominates.

The cross section upper limits are obtained neglecting the contribution of
direct $\stone$ squark pair production, which can satisfy the
selection criteria for the single-lepton or OS-lepton SRs if a
light-parton jet is misidentified as a $\cPqb$ jet or if there is additional radiation leading to genuine $\cPqb$ jets.
Including direct $\stone$ squark pair production in the single-lepton
or two OS lepton SRs typically lowers the cross section limit by
a few percent, with the most pronounced differences occurring at larger
$\sttwo$ mass. The contribution in
the case of events with two SS leptons or at least three leptons is
small due to the low probability of misidentifying non-prompt leptons.
Since the signature with three leptons has the best sensitivity
overall, the impact on the combined limit is much smaller than the
uncertainty in the production cross section.

\section{Summary}
\label{sec:summary}

This letter presents results of a search for the pair production of the
heavier top-squark mass eigenstate $\sttwo$ decaying to the lighter
eigenstate $\stone$, producing a signature of a top-antitop quark pair
in association with Higgs or $\cPZ$ bosons.
The analysis explores final states with exactly one lepton and at least three identified bottom-quark jets (\bjets),
with exactly two leptons of opposite charge and at least three \bjets, with exactly two same-sign leptons and at least one \bjet,
and with three or more leptons and at least one \bjet, where by ``lepton'' we mean an electron or muon.
No significant excess event yield above standard model expectations is
observed. The results are used to exclude a range of $\sttwo$ masses
below approximately 575\GeV for $\stone$ masses below approximately
400\GeV. The interpretation assumes that the $\stone$ squark always decays to $\cPqt\PSGczDo$ and
that $m_{\stone} - m_{\PSGczDo} \simeq m_{\cPqt}$, where the $\PSGczDo$ particle represents a stable,
weakly interacting lightest supersymmetric particle neutralino whose signature in the detector is missing transverse energy. This region of phase space is not probed by searches for direct $\stone$ squark pair production.

\section*{Acknowledgements}

We congratulate our colleagues in the CERN accelerator departments for the excellent performance of the LHC and thank the technical and administrative staffs at CERN and at other CMS institutes for their contributions to the success of the CMS effort. In addition, we gratefully acknowledge the computing centers and personnel of the Worldwide LHC Computing Grid for delivering so effectively the computing infrastructure essential to our analyses. Finally, we acknowledge the enduring support for the construction and operation of the LHC and the CMS detector provided by the following funding agencies: BMWFW and FWF (Austria); FNRS and FWO (Belgium); CNPq, CAPES, FAPERJ, and FAPESP (Brazil); MES (Bulgaria); CERN; CAS, MoST, and NSFC (China); COLCIENCIAS (Colombia); MSES and CSF (Croatia); RPF (Cyprus); MoER, ERC IUT and ERDF (Estonia); Academy of Finland, MEC, and HIP (Finland); CEA and CNRS/IN2P3 (France); BMBF, DFG, and HGF (Germany); GSRT (Greece); OTKA and NIH (Hungary); DAE and DST (India); IPM (Iran); SFI (Ireland); INFN (Italy); NRF and WCU (Republic of Korea); LAS (Lithuania); MOE and UM (Malaysia); CINVESTAV, CONACYT, SEP, and UASLP-FAI (Mexico); MBIE (New Zealand); PAEC (Pakistan); MSHE and NSC (Poland); FCT (Portugal); JINR (Dubna); MON, RosAtom, RAS and RFBR (Russia); MESTD (Serbia); SEIDI and CPAN (Spain); Swiss Funding Agencies (Switzerland); MST (Taipei); ThEPCenter, IPST, STAR and NSTDA (Thailand); TUBITAK and TAEK (Turkey); NASU and SFFR (Ukraine); STFC (United Kingdom); DOE and NSF (USA).

Individuals have received support from the Marie-Curie programme and the European Research Council and EPLANET (European Union); the Leventis Foundation; the Alfred P. Sloan Foundation; the Alexander von Humboldt Foundation; the Belgian Federal Science Policy Office; the Fonds pour la Formation \`a la Recherche dans l'Industrie et dans l'Agriculture (FRIA-Belgium); the Agentschap voor Innovatie door Wetenschap en Technologie (IWT-Belgium); the Ministry of Education, Youth and Sports (MEYS) of the Czech Republic; the Council of Science and Industrial Research, India; the Compagnia di San Paolo (Torino); the HOMING PLUS programme of Foundation For Polish Science, cofinanced by EU, Regional Development Fund; and the Thalis and Aristeia programmes cofinanced by EU-ESF and the Greek NSRF.

\bibliography{auto_generated}

\cleardoublepage \appendix\section{The CMS Collaboration \label{app:collab}}\begin{sloppypar}\hyphenpenalty=5000\widowpenalty=500\clubpenalty=5000\textbf{Yerevan Physics Institute,  Yerevan,  Armenia}\\*[0pt]
V.~Khachatryan, A.M.~Sirunyan, A.~Tumasyan
\vskip\cmsinstskip
\textbf{Institut f\"{u}r Hochenergiephysik der OeAW,  Wien,  Austria}\\*[0pt]
W.~Adam, T.~Bergauer, M.~Dragicevic, J.~Er\"{o}, C.~Fabjan\cmsAuthorMark{1}, M.~Friedl, R.~Fr\"{u}hwirth\cmsAuthorMark{1}, V.M.~Ghete, C.~Hartl, N.~H\"{o}rmann, J.~Hrubec, M.~Jeitler\cmsAuthorMark{1}, W.~Kiesenhofer, V.~Kn\"{u}nz, M.~Krammer\cmsAuthorMark{1}, I.~Kr\"{a}tschmer, D.~Liko, I.~Mikulec, D.~Rabady\cmsAuthorMark{2}, B.~Rahbaran, H.~Rohringer, R.~Sch\"{o}fbeck, J.~Strauss, A.~Taurok, W.~Treberer-Treberspurg, W.~Waltenberger, C.-E.~Wulz\cmsAuthorMark{1}
% ...
\vskip\cmsinstskip
% ...
1:~~Also at Vienna University of Technology, Vienna, Austria\\
2:~~Also at CERN, European Organization for Nuclear Research, Geneva, Switzerland\\
% ...

\end{sloppypar}
\end{document}